%% file: masterfile.tex
\documentclass[12pt, a4paper]{article}

\usepackage{dsfont}
\usepackage{fullpage,graphicx}
\usepackage{amsmath,amssymb,amsfonts,amsthm,mathtools}
\usepackage{natbib,float,enumerate,tabularx,booktabs,url,comment,enumitem}
\usepackage{geometry}
\usepackage[onehalfspacing]{setspace}
\usepackage[titletoc,toc]{appendix}
\usepackage{minitoc}
\renewcommand \thepart{}
\renewcommand \partname{}
\usepackage[flushleft]{threeparttable}
\usepackage{ragged2e}
\usepackage{caption}
\usepackage{subcaption}
  \usepackage{tikz}
		\usetikzlibrary{chains, positioning, arrows}
\usepackage{lscape}
\usepackage{setspace}
\usepackage[parfill]{parskip}        % Activate to begin paragraphs with an empty line rather than an indent
\usepackage{xcolor} % For color support

\usepackage[thinc]{esdiff}
\usepackage{bm,bbm}

\newtheoremstyle{break}
  {\topsep}{\topsep}%
  {\itshape}{}%
  {\bfseries}{}%
  {\newline}{}%
\theoremstyle{break}

\newtheoremstyle{proposition}
  {\topsep}{\topsep}%
  {\itshape}{}%
  {\bfseries}{}%
  {\newline}{}%
\theoremstyle{proposition}

\theoremstyle{definition}

        \renewcommand\tilde[1]{\widetilde{#1}}

\usepackage{tocvsec2}
\usepackage{xpatch}
\makeatletter
\xpatchcmd{\sv@part}{\huge \bfseries \partname \nobreakspace \thepart \par \vskip 20\p@ \fi \Huge \bfseries #2}{\fi \Huge \bfseries \thepart. #2}{}{}
\makeatother

\renewcommand \thepart{}
\renewcommand \partname{}

\begin{document}
\onehalfspacing
\doparttoc % Tell to minitoc to generate a toc for the parts
\faketableofcontents % Run a fake tableofcontents command for the partocs

%\part{} % Start the document part

\def \thetitle{The Cost of Delivery Delays}
	 \title{The Cost of Delivery Delays\footnote{This paper was prepared for the 2025 AEA Papers and Proceedings. We are thankful to Ryan Kim for useful feedback. Ferrari gratefully acknowledges financial support from the Swiss National Science Foundation (grant number 100018-215543). Yansong Zhang and Linen Yu provided excellent research assistance.}}% Full title goes here	
	\def \theshorttitle {} %Short title goes here
	\author{ \color{blue}\large{Maria Jose Carreras-Valle}\color{black}
\\ \small \hspace{-8pt} Pennsylvania State University \and
\color{blue}\large{Alessandro Ferrari}\color{black}
\\ \small \hspace{-8pt} Universitat Pompeu Fabra, BSE \& CEPR
}
	\def \thedate {\today} 

\maketitle
\vspace{-15pt}
\begin{abstract}

\input{draft/0_abstract}

\end{abstract}

\begin{raggedright} Keywords: Inventories, tariffs, delivery times, delays, supply risk, supply pressures.\\
%JEL Codes: \\
\end{raggedright}
\newpage

\onehalfspacing
\input{draft/1_intro}
\input{draft/2_data}

\input{draft/3_model}

\input{draft/4_quantification}
\input{draft/4_results}

\input{draft/5_conclusions}

\center{\addcontentsline{toc}{section}{ References} }
\bibliographystyle{aer}
{\footnotesize\bibliography{input/lit}}
\begin{appendices}
\input{draft/6_appendix}

\end{appendices}

\end{document}

%% file: draft/0_abstract.tex
Since 2018, there has been a consistent decline in the distance traveled by U.S. manufacturing imports, reaching a level not observed since 2008. This trend is the result of the substitution away from imports from China and towards imports from closer countries. At the same time, U.S. manufacturing inventory-to-sales ratio has continued to rise. These trends are at odds with the literature, which finds that reductions in the distance of imports are associated with a decline in inventories. We argue that a rise in delivery time risk, driven by longer and more frequent delays and supply disruptions, can reconcile these trends. We do so in the context of a model of global sourcing with stochastic delivery times and inventories. Firms trade off the lower price of farther inputs with the increase in exposure to demand volatility and longer delays. In response, firms increase their inventories. Yet, as delivery delays rise, firms need to carry more inventories per unit of the input used. We calibrate the model for the period from 2018 to 2024 using data on the increase in tariffs for inputs from China, and the rise in inventories over sales. We find an increase in delivery delays for foreign inputs of 21 days across the period. The rise in delays and tariffs had an output loss of $7.3\%$ and a price increase of $1.8\%$. Of these, the rise in delivery delays alone generated a $2.6\%$ drop in output and a $0.4\%$ increase in prices.

%% file: draft/1_intro.tex
\section{Introduction}

Modern economies organize production along complex supply chains, whose operation relies on the smooth functioning of logistics. Recent events have strained the international transport system and generated frequent supply disruptions. Yet, measures of supply pressures or delays are surprisingly difficult to estimate. 

In this paper, we provide a model-based quantification of delays in supply procurement and study their social cost. We consider an economy in which firms face demand volatility and their foreign inputs face positive and stochastic delivery times. Firms require inputs from China, rest of the world (ROW), and domestic to produce. For the inputs from China and ROW, firms have access to a fraction of the order they placed in the period, and the rest of the order firms can't use to produce. Further, the share of the arriving order is stochastic. Thus, firms hold inventories of their inputs due to the interaction between the demand shocks and the positive delivery times, and additionally due to the delays inputs may face. 

We rely on two empirical observations to quantify our model. First, the average import distance of U.S. imports has started declining after decades of steady increase. The decline in distance is driven by the decrease in imports from China, due to the rise in tariffs applied in 2018. Second, the inventory-to-sales ratio of U.S. manufacturing firms has stabilized at a permanently higher level after the global pandemic in 2020. These data trends are surprising, considering the well-established positive correlation between the distance imports travel and inventories stocked, as documented in the literature. We argue that a rise in delivery risk, driven by longer and more frequent delays and supply disruptions, can reconcile these trends. The National Association of Manufacturers (NAM) Manufacturers Outlook Survey for 4th quarter of 2023 reports that $86.2\%$ of the firms in their sample report have worked to `de-risk' their supply chains in the last two years. 

We use the model to obtain a measure of delays that is able to rationalize both trends. We start by calibrating the model to match moments of the U.S. manufacturing sector in 2018. Then, we use data on the rise in tariffs applied by the Trump and Biden administrations to inputs from China, the decline in the imports from China, and the rise in input inventories from 2018 to 2024 to calibrate the trade war with China that started in 2018, and create the measure for the rise in delivery delays, or a rise in  in the standard deviation of the distribution of delivery days, for foreign inputs.

Using the increase in tariffs between 2018 and 2019, we show that the import substitution from Chinese to the ROW inputs can rationalize the declining import distance. However, absent any other structural change, this would imply a reduction in the stock of inventories held by U.S. firms since the average delay for Chinese imports is estimated to be larger than that of ROW imports. Therefore, we consider an increase in delivery delays for all imported inputs. We use our model to estimate this increase since 2018. The average delay faced by U.S. importers was between 10 and 5 days when importing from China and ROW in 2018, respectively. These became 31 and 26 days at the onset of the global pandemic in 2020 and, after a short decline, are peaking again in 2024.

We compare our model based measure of delivery delays to the private data on lead times reported by the Institute of Supply Management (ISM) used in their Manufacturing PMI Index. The lead times for maintenance, repair, and operating supplies (MRO), production materials, and capital expenditures observe a sharp increase in 2020 and in 2024 remain higher than pre-Covid levels. Using our estimated distributions of delivery days, and the calibrated rise in delays, we compute a measure of days using a similar methodology as ISM. We find our measure of delays is similar to the ISM data. MRO supplies increase by a total of 11 days and production materials by 14 days, while our measure shows an increase of 11 days.

We then compute the implications of the rise in delivery delays in terms of output and prices. First, we find that the combination of tariffs and delivery delays increase generated an output drop of 7.3\% and an increase in prices of 1.8\%. These effects can be decomposed through the lens of the model. We estimate that the trade war alone induced a 5.1\% drop in output and a 1.4\% increase in prices, while the rising delivery delays alone would have generated a 2.6\% drop in output and 0.4\% increase in prices.

\textbf{Related Literature}

%At the same time, geopolitical tensions have further hindered free trade. Notably, in 2018, the Trump administration levied tariffs on Chinese goods, which the Chinese government retaliated against. 
This paper contributes to several strands of literature. First, we contribute to a large body of work that analyzes the effect of disruptions in the production process, as in \citet{barrot2016}, \citet{carva2020},  \citet{AKK2020}, \citet{akk2021}, \citet{Alessandriaetal2023}, \citet{acemoglu2024macroeconomics}. This paper is closest to \citet{Alessandriaetal2023}, where they quantify the costs of disruptions during COVID-19. Our contribution to this literature is to provide a measure of disruptions through delivery delays using public inventory and import data. 

We incorporate the rise in tariffs between the U.S. and China that started in 2018. We rely on the tariff estimate provided by \citet{amiti} and follow the discussion on \citet{faj2021} and \citet{caliendo2023}. Here, we use a model with foreign trade and inventories to isolate and quantify the cost of the rise in tariffs from the cost of disruptions that were also present from 2018 to 2024. 

Third, there is a strand of work that studies the importance of delivery times for trade, as in \citet{Hummels2007}, 
\citet{HummelsSchaur2013}, \citet{clark2024}, and \citet{baum2024}. Our contribution to this literature is to provide a measure of delivery delays and compare it to the private data on lead times. Our measure matches well with the data on lead times, and we use the model to quantify the cost of the delivery times in terms of output and prices.

Last, this paper touches on the large literature on inventory management, as in the works by \citet{KhanThomas2008}, \citet{AKM2010}, \citet{KhanKheder2020_2}, \citet{Pish2020}, \citet{carreras2021increasing}, \citet{ferrari2022inventories}. The model is based on \citet{carreras2021increasing}, where delivery time friction for foreign inputs is introduced as the firm is able to use only a share of the order today, and the rest is delivered until the next period. Here, we leverage the model to introduce our measure of delivery delays.

The remainder of the paper proceeds as follows. Section 2 describes the motivating evidence and the data used. Section 3 outlines the model and discusses the trade-off between the price and delivery times across input suppliers. Section 4 presents the calibration of the model used to perform our quantitative analysis in Section 5. Here, we present our model-based measure of delays and the costs of the rise in tariffs and delays. Section 6 concludes. 

%% file: draft/2_data.tex
\section{Motivating Evidence}
\begin{figure}
\centering
\caption{Rise in delays: distance fell while inventories remain high}
\begin{subfigure}{.475\textwidth}
  \centering
  \includegraphics[width=\textwidth]{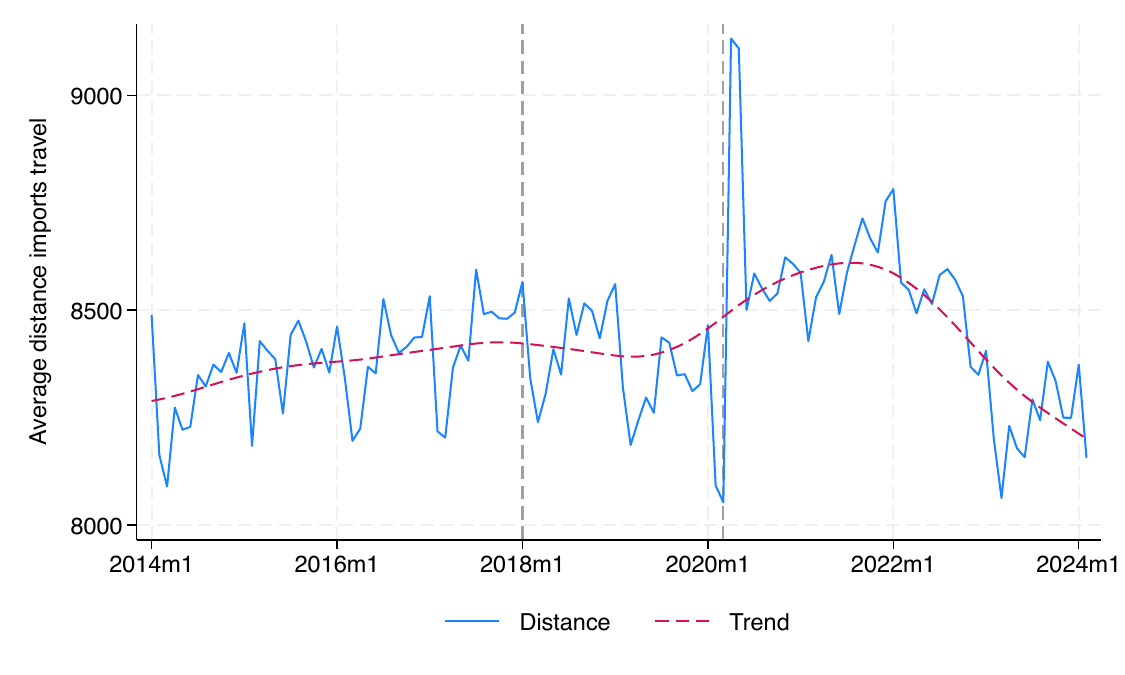}
  \caption{Distance traveled by manufacturing imports}
\end{subfigure}%
\begin{subfigure}{.475\textwidth}
  \centering
  \includegraphics[width=\textwidth]{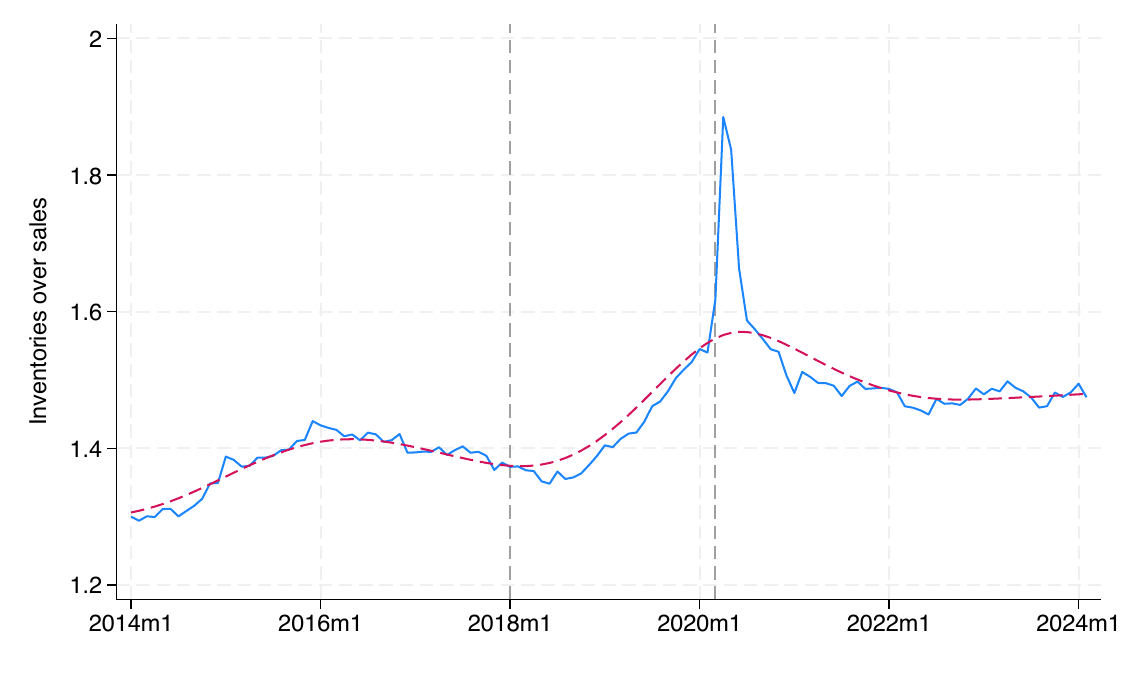}
  \caption{Rise in manufacturing inventories over sales }
\end{subfigure}
\label{fig:main}
\end{figure}

Our research question is motivated by two empirical observations shown in Figure \ref{fig:main}. First, we show that the average distance U.S. imports travel has been decreasing, and in 2024, it reached the lowest point since 2014.% The decrease in the distance imports travel is driven by the decline in imports from China, that starts with the tariff hikes in 2018.
The second empirical observation is that U.S. manufacturing inventories to sales ratio have increased from 2018 to 2024. The rise is observed across manufacturing sectors and types of inventories. %We relate these observations to the behavior of delays, disruptions, and pressures firms face during this period.  and highlight their limitations to estimate and quantify the rise in volatility in the post-Covid period.

First, we follow \cite{ganapati2023far} and build a measure of the distance traveled by U.S. imports. The index in equation \ref{eq:distance} combines information from the CEPII-BACI dataset on the average distance between countries and monthly import data from the U.S. Census Bureau by country of origin.
\begin{align}  \label{eq:distance}
    D_t = \sum_i \frac{D_{i} M_{it}}{\sum_i M_{it}},
\end{align}
where $D_{it}$ is the distance between country $i$ and the U.S. and $M_{it}$ is U.S. manufacturing imports from $i$ at time $t$.

\begin{figure}
\centering
\caption{Decline in manufacturing imports from China since 2018}
\begin{subfigure}{.475\textwidth}
  \centering
  \includegraphics[width=\textwidth]{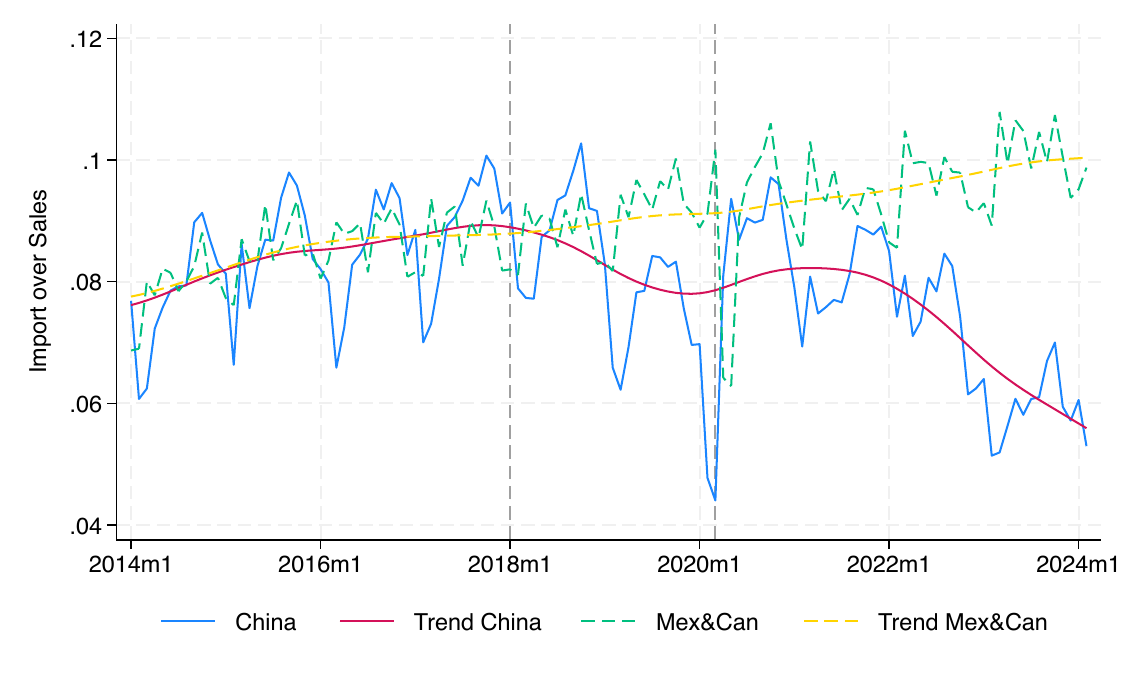}
  \caption{Decline in the imports from China}
\end{subfigure}%
\begin{subfigure}{.475\textwidth}
  \centering
  \includegraphics[width=\textwidth]{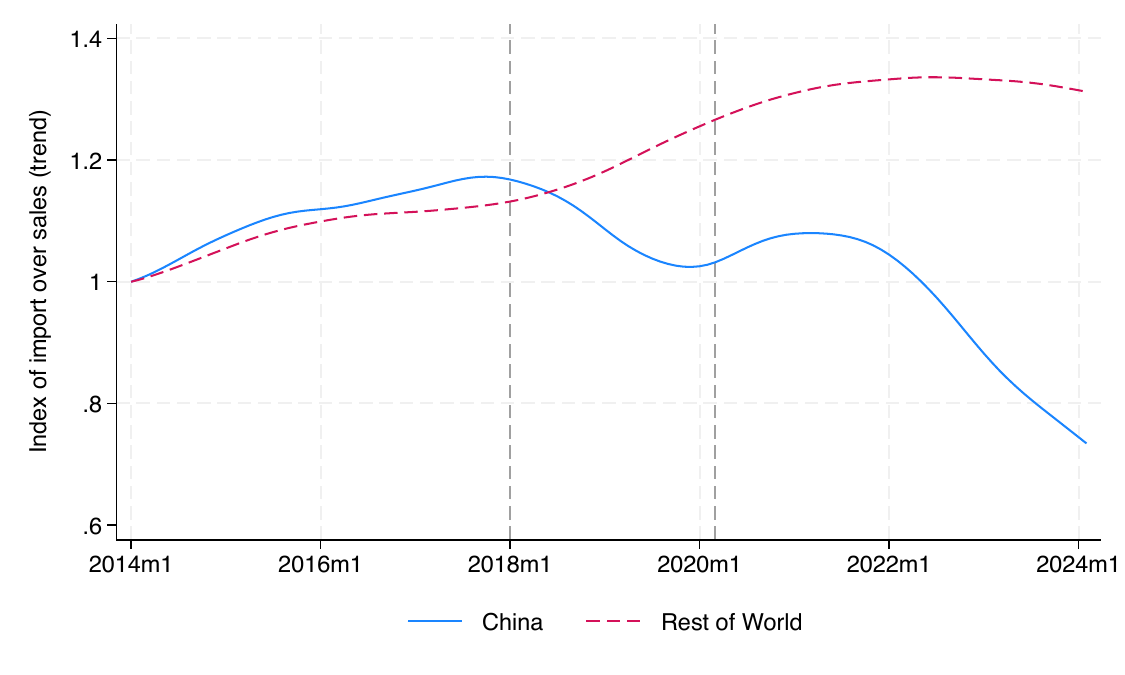}
  \caption{Imports from China and ROW over sales }
\end{subfigure}
\label{fig:china1}
\end{figure}

\begin{figure}
\centering
\caption{Rise in U.S. tariffs on imports from China}
\begin{subfigure}{.475\textwidth}
  \centering
  \includegraphics[width=\textwidth]{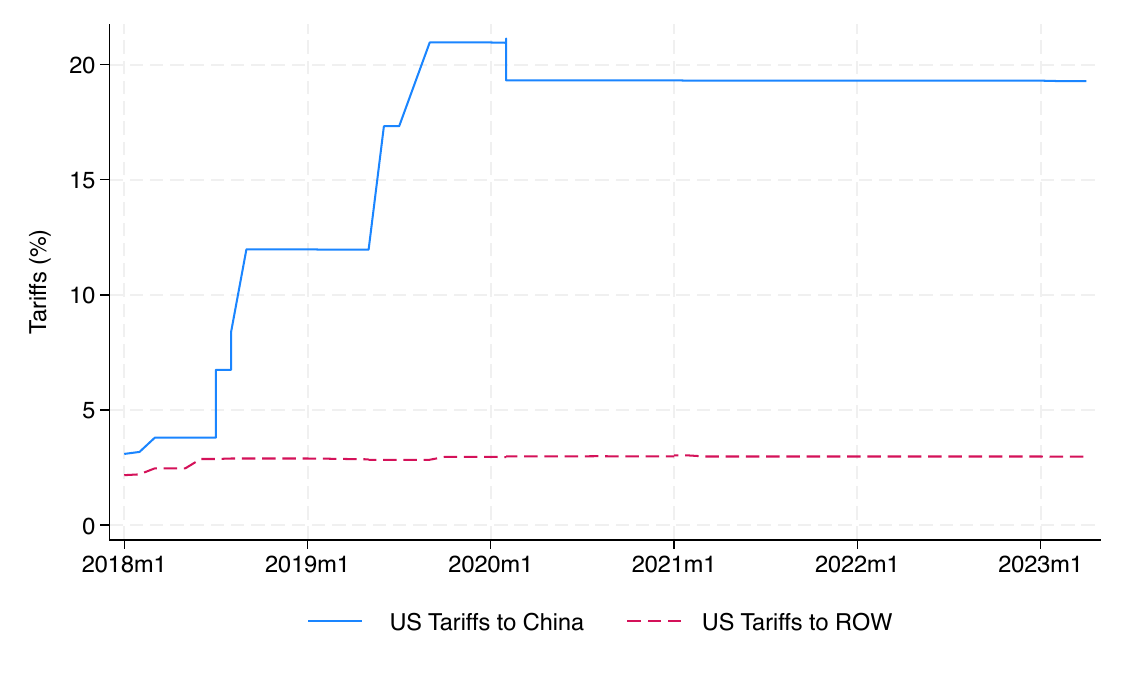}
  \caption{Rise in tariffs starts in 2018}
\end{subfigure}%
\begin{subfigure}{.475\textwidth}
  \centering
  \includegraphics[width=\textwidth]{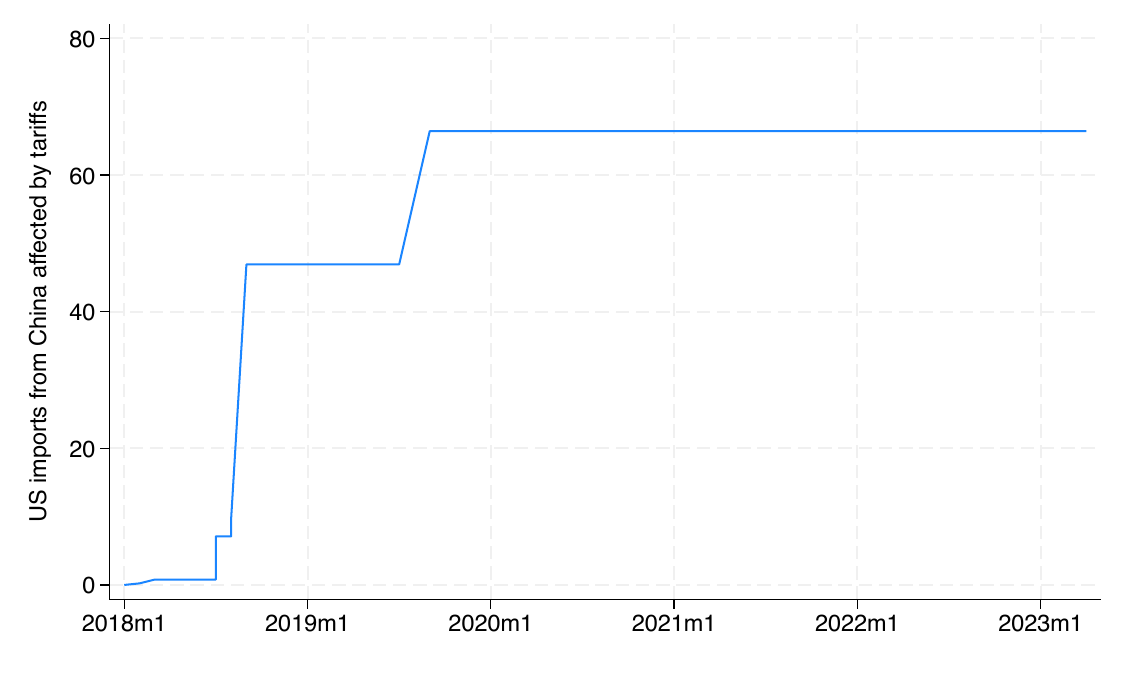}
  \caption{Affects the majority of products imported from China }
\end{subfigure}
\label{fig:tariff}
\end{figure}

The drop in import travel distance is driven by the decrease in the use of Chinese inputs, as shown in panel (a) in Figure \ref{fig:china1}. Here, we show monthly manufacturing imports over sales for the three main U.S. trade partners (China, Mexico, and Canada) and their HP-filtered trends. The trend for the imports from China over sales declined from $8.9\%$ in 2018 to $5.7\%$ in 2024, whereas the sum of the imports from Mexico and Canada steadily grew between 2014 and 2024. Panel (b) compares the growth of the imports from China over sales to the imports from the rest of the world (ROW). Around 2018, we observe a trend reversal on Chinese imports while imports from the ROW remain on trend. We interpret this as a consequence of substitution away from Chinese imported inputs. Both of these trends are consistent across NAICS three-digit manufacturing sectors, as shown in Appendix \ref{appendix:data}.
The decline in imports from China follows the tariff hikes imposed by the U.S. government, which started in 2018. Using data from the Peterson Institute for International Economics (PIIE), Panel (a) in Figure \ref{fig:tariff} shows the increase in U.S. tariffs to China from 2018 to 2020, while tariffs to the ROW remained relatively constant. Further, Panel (b) shows that the majority of products imported from China were affected, and by 2020 around $65\%$ of imports from China faced an increase in tariff. Following \citet{amiti}, we compute the average tariff change for imports from China using tariff data from the U.S. International Trade Commission (USITC) and U.S. Trade Representative (USTR), and import flows from the U.S. Census Bureau as follows
\begin{align}
    \Delta \tau =\frac{\sum_i \Delta \tau_{i, 2020 - 2017} M_{i, 2017}}{\sum_i M_i, 2017}, 
\end{align}
with $\Delta \tau$ is the weighted average change in tariff between 2017 and 2020, and $M_{i, 2017}$ is the U.S. product $i$ import value from China in January 2017. Table \ref{table:tariffs} shows the tariff-weighted average increase for imports from China of $15$ percentage points.

\begin{table}[h!]
\centering
\caption{Increase in U.S. tariffs for imports from China}
\begin{tabular}{@{}lcc@{}}
\textbf{Tariff Rate} & \textbf{Weighted Average} & \textbf{Average} \\ \midrule
2017m1 & 2.99\% & 4.82\% \\
%2019m1 & \textbf{0.09203777} & 0.1341093 \\
%$\Delta$ (2017m1-2019m1) & \textbf{0.06206906} & 0.085879 \\
2020m1 & 18.05\% & 24.49\% \\
$\Delta$ (2017m1-2020m1) & {15.05\%} &19.66\% \\
\end{tabular}
\label{table:tariffs}
\end{table}

\begin{figure}
\centering
\caption{Rise in manufacturing inventories to sales}
\begin{subfigure}{.475\textwidth}
  \centering
  \includegraphics[width=\textwidth]{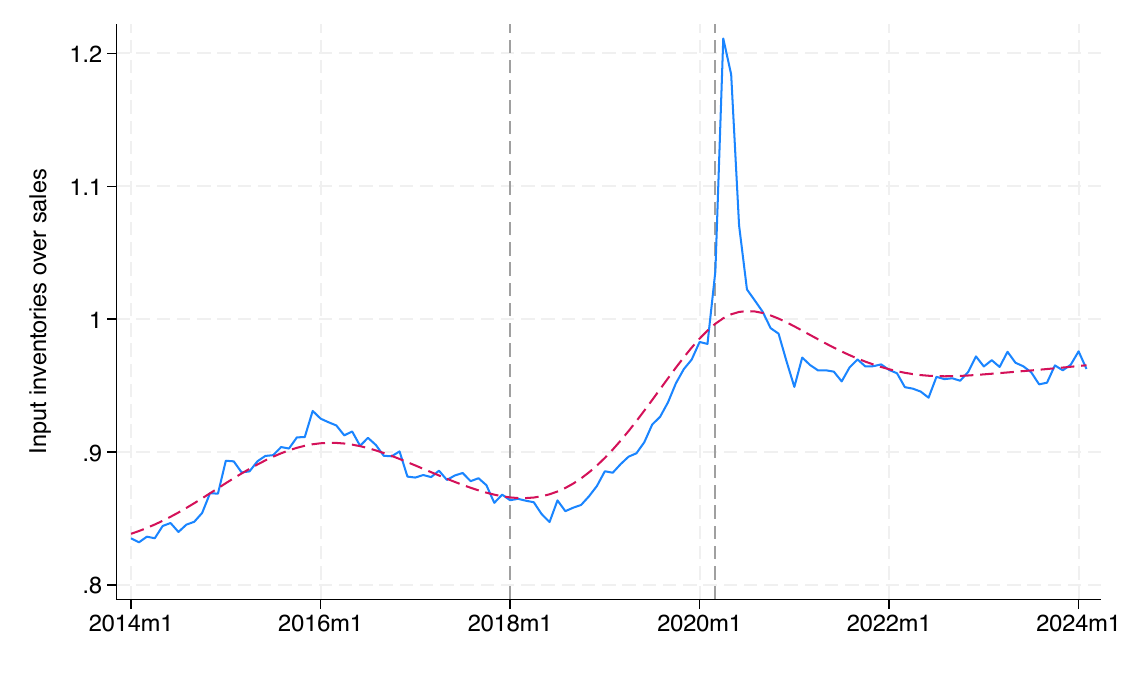}
  \caption{Rise in input inventories over sales}
\end{subfigure}%
\begin{subfigure}{.475\textwidth}
  \centering
  \includegraphics[width=\textwidth]{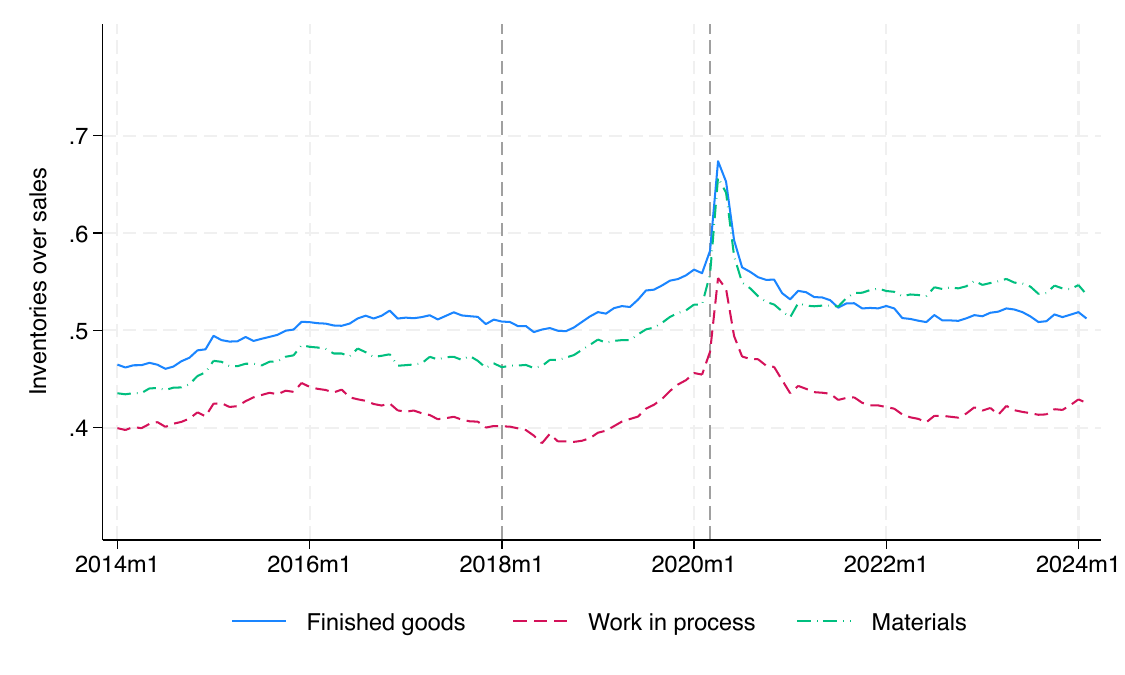}
  \caption{Rise across types of inventories }
\end{subfigure}
\label{fig:inv2}
\end{figure}

As the distance imports travel declines, we expect the mean of the delivery times of inputs for U.S. firms to decrease. If firms hold inventories to ensure smooth production operations, we should also expect lower inventory stocks since their inputs are more readily available. However, we observe an increase in the U.S. manufacturing inventories over sales from 2018 to 2024. Figure \ref{fig:inv2} shows the rise in input inventories over sales, measured as the sum of material and supplies and work-in-process inventories, reported by the U.S. Census Bureau. In 2018, the average firm carried around $26$ days of sales as input inventories, and by 2024, they carry an additional 3 days of sales in input inventories. Panel (b) shows the rise across the three types of inventories: finished goods, materials and supplies, and work-in-process inventories. The sharpest rise occurs in the materials and supplies inventories and then in work-in-process inventories. Finished good inventories remain relatively constant, yet neither shows the decrease expected given the decrease in the mean of delivery times. The rise in inventories in this period is observed across NAICS three-digit manufacturing sectors, as shown in Appendix \ref{appendix:data}.

\begin{figure}
\centering
\caption{Rise in disruptions and lead times}
\begin{subfigure}{.475\textwidth}
  \centering
  \includegraphics[width=\textwidth]{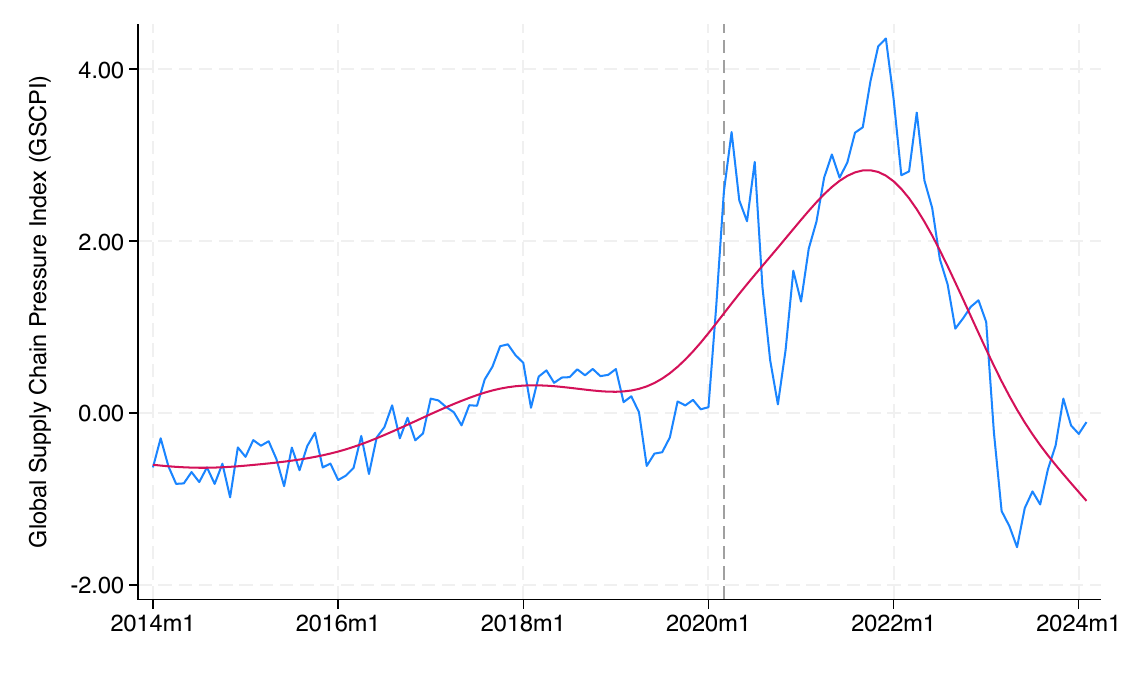}
  \caption{Global Supply Chain Pressure Index}
\end{subfigure}%
\begin{subfigure}{.475\textwidth}
  \centering
  \includegraphics[width=\textwidth]{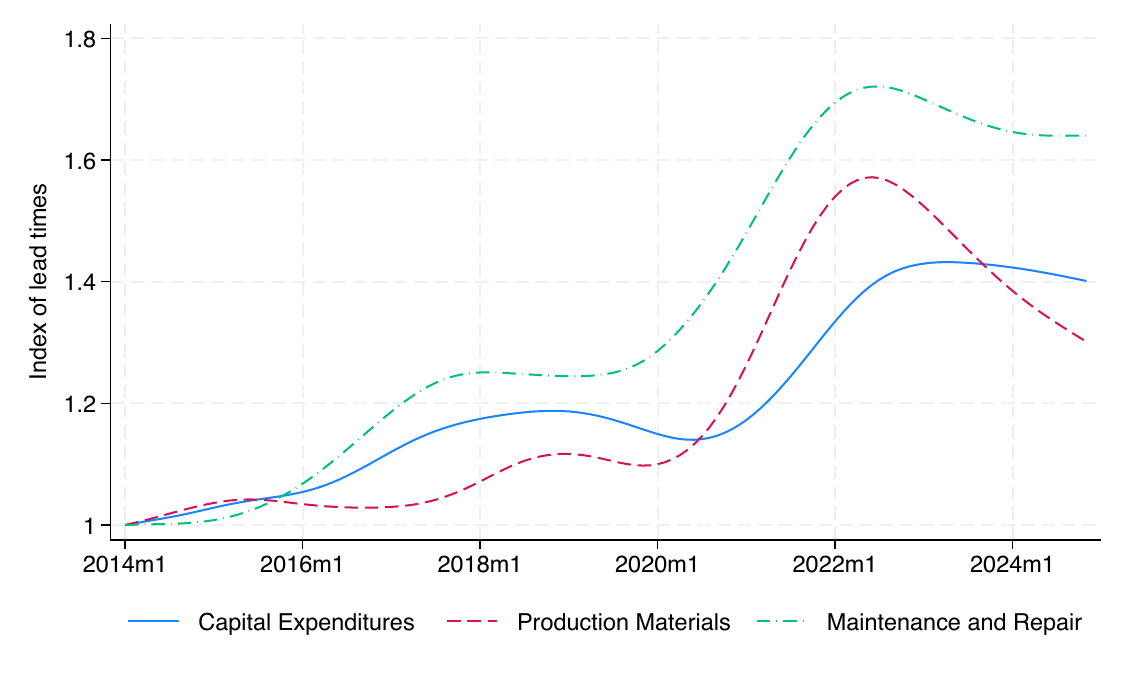}
  \caption{Rise in the lead times reported by the ISM}
\end{subfigure}
\label{fig:pressure}
\end{figure}

These seemingly conflicting observations can be reconciled by rising delays and disruptions in supply chains. To investigate this hypothesis, we present two measures that show evidence of the supply chains pressure firms faced during and after Covid-19. The first measure is the Global Supply Chain Pressure Index (GSCPI) reported by the Federal Reserve Bank of New York, shown in Panel (a) in Figure \ref{fig:pressure}. The GSCPI aims to provide evidence of supply pressures faced by U.S. firms, and shows an increase starting in 2019, which is reabsorbed by 2023. Importantly, this does not imply that the pressures firms face are back to their 2018 level. The index uses data on transportation costs and supply-chain components. They combine data on the cost of container shipping and the charter market, reported by the Baltic Dry Index (BDI) and the Harper Index, and air freight rates reported by the Bureau of Labor Statistics, which is $37\%$ higher in 2024 than in 2018. Appendix \ref{appendix:data} shows the data on transportation costs. The data on supply-chain components, provided by the Institute of Supply Management (ISM) Purchase Manager Index (PMI), reports diffusion indices on new orders, production, employment, and deliveries of manufacturing firms. Diffusion indices measure changes in the extensive margin, the number of firms that report an increase or decrease, but not the change in the intensive margin. As a consequence, they do not reflect the changes in the level of pressures or disruptions firms face. Thus, we cannot use the GSCPI to measure and quantify the volatility firms face, nor are we able to compare the environments across time. 

Next, the ISM has private data on average lead times for inputs for manufacturing firms, which have been consistently rising since 2020. The lead times are computed by measuring the number of firms that report their lead time in different bins of time, for 5, 30, 60, 90, 180, or 365 days or more. Panel (b) of Figure \ref{fig:pressure} shows the index of the growth of lead times for three types of inputs, capital expenditures, maintenance, repair and supplies, and production materials. After 2022, lead times decreased or remain constant, yet the level is still higher than the pre-Covid 2018 levels. As the mean of delivery times decline, given the drop in the distance imports travel, the rise in lead times shows evidence of a rise in delays. However, this observational data naturally includes any optimal rerouting by firms as a consequence of rising delays. Therefore, we cannot use these lead times directly as the shock. In what follows, we use our model to back out the rise in delays accounting for the optimal choices of firms. We do so only using the model and public data on U.S. manufacturing inventories and compare our measure to the ISM data. Further, the model will allow us to isolate and quantify the costs associated with the rise in delays and tariffs. 

%\textcolor{red}{However, with these data series one cannot disentangle and measure the effects of the mean vs the variance of delivery times. To overcome this limitation, in the next section, we build a model to quantify the rise and costs of delays using data on imports and inventories.} With our lognormal assumtpion I am not sure we can do this.

%% file: draft/3_model.tex
\section{Model}

We present a model with stochastic delivery times for inputs and demand volatility to measure the rise in delays and quantify their cost in terms of output, prices, and sourcing choices. Following the framework in \citet{carreras2021increasing}, we introduce a model where firms use three different inputs to produce a final good. They require a domestic input, a foreign input sourced from a faraway trade partner, which represents trade with China, and a foreign input from a close trade partner, which stands for trade with the rest of the world (ROW). We assume domestic inputs are delivered within the period, while both foreign inputs face pair-specific and stochastic delivery times. Firms are able to stock inventories of foreign inputs to insure against demand and delivery time shocks.

\subsection{Environment}
Time is discrete and indexed by $t \in \{0,1,2, ..., \infty \}$. The partial equilibrium economy is composed of a unit continuum of monopolistically competitive final good producers, a unit continuum of firms that produce a domestic input, two foreign inputs, and a domestic representative consumer.  Uncertainty in the model is given by firm-specific, independent, and identically distributed (\textit{iid}) demand shocks that each firm faces from the consumer and delivery time shocks for each of the foreign inputs.

%%%%%
\textbf{Final good firms.}
The unit continuum of final good firms, $j \in [0,1]$, behave monopolistically and produce a unique variety of the final good, $y_j$. Final good firms maximize profits subject to six constraints. First, each firm faces the demand from the representative consumer, which is a function of the per-period $iid$ firm-specific demand shock, $\nu_j$, total production, $Y$, and price index, $P$.
\begin{equation} 
\label{demand}
y_j(p_j) \: = {\nu_j} \: Y \: (P/p_j)^{\epsilon} \: \: \: \: \: \: \: \: \: \: \: \: \: \: \: \: \: \: \text{where} \: \: \nu_{j} \sim_{iid} G\:(\mu_{\nu}, \: \sigma_{\nu}) 
\end{equation}  	
 	
Second, a firm's technology combines the domestic input, $x_j^d$, a foreign input sourced from a close trade partner, $x_j^c$, and foreign input sourced from a far trade partner, $x_j^f$, to produce the final good. I assume the inputs are combined using a constant elasticity of substitution (CES), with elasticity $\sigma$. Within the CES aggregator, close foreign inputs are weighted using $\theta^c$ and the far foreign inputs using $\theta^f$. This allows the model to match the initial level of inputs observed in the data. 
\begin{equation} 
\label{tech}
y_j = \Big( \:  {\theta^c}^{ \frac{1}{\sigma}} \: x_j^{c \: \frac{\sigma -1}{\sigma}}  \: + \: {\theta^f}^{\frac{1}{\sigma}} \: x_j^{f \: \frac{\sigma -1}{\sigma}} \: + \: (1 - \theta^c - \theta^f)^{\frac{1}{\sigma}} \: x_j^{d \: \frac{\sigma -1}{\sigma}}  \Big)^{\frac{\sigma}{\sigma-1} }
\end{equation}  

Third, the close and far foreign inputs, $i = \{c,f\}$, face stochastic delivery times. Only a fraction $\lambda^i_j$ of the order of the inputs, $n^i_j$, is available for them to produce that period. Thus, inputs used to produce, $x_j^i$, are constrained to be less than or equal to the initial level of inventories, $s_j^i$, and the fraction $\lambda_j^i$ of the order that arrives before production takes place. Each delivery time shock, $\lambda_j^i$, is drawn from the input-specific distribution $F^i(\cdot)$. Equation (\ref{inputs}) summarizes the third and fourth constraint the final good firm faces. Fourth, firms are able to store inventories of each foreign input, which depreciate at rate $\delta$. Equation \ref{lominv} shows the law of motion of inventories, in which inventories tomorrow, $s_j^{\prime i}$, are equal to the inputs left after production, $s_j^i \: + \: \lambda_j^i \: n_j^i \: - \: x_j^i$, plus the remaining order arriving next period, $(1- \lambda_j^i) \: n_j^i$, discounted at rate $(1-\delta)$. We assume that, on average, the close foreign input face shorter delivery times than the far foreign input, $E(\lambda^c) > E(\lambda^f)$. The domestic input, $x_j^d$, can be delivered entirely within the period, so the firm has no incentives to hold inventories for this input.
\begin{equation}
\label{inputs}
\begin{split}
  x_j^i \:  \leq  \: s_j^i \: + \: {\lambda_j^i} \: n_j^i   \: \:  \: \: \: \:  \: \: \: \: \: \: \:  \: \: \: \: \:  \:  \: \: \: \: \:  \: \: \: \: \:  \:  
 \text{where} \: \: \: \lambda^i_{j} \sim_{iid} \: F^i(\mu_{\lambda}^i, \sigma_{\lambda}^i)       
  \end{split}
\end{equation}
  \begin{equation}
\label{lominv}
\begin{split}
  s_j^{\prime i} = (s_j^i \: + \:n_j^i \: - \: x_j^i)  (1\: - \: \delta) \:   \:  \: \: \: \: \:  \: \: \: \: \:  \: \: \: \: \:  \: \: \: \: \:  \: & \: \: \:   \text{for} \: \: \: i = \{c,\: f \} 
\end{split}
\end{equation}

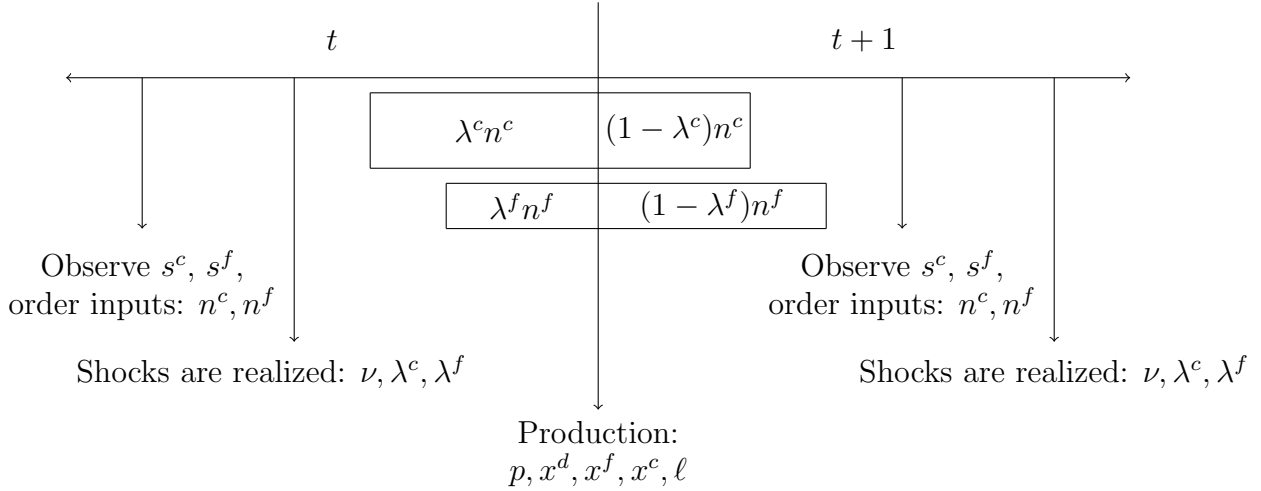
\begin{figure}
\centering
\caption{Timing -- firms order inputs before shocks are realized}
\begin{tikzpicture}

\draw[<->] (0,0) -- (14,0);

\draw[->] (1,0) -- (1,-2);
\draw (1,-2.5) node{Observe $s^c$, $s^f$,};
\draw (1,-3) node{order inputs: $n^c, n^f$};

\draw[->] (3,0) -- (3,-3.5);
\draw (2.7,-3.9) node{Shocks are realized: $\nu, \lambda^c, \lambda^f$};

\draw [->] (7,0) -- (7,-4.4);
\draw (7,-4.7) node{Production: };
\draw (7,-5.2) node{ $p, x^d, x^f, x^c, \ell$ };

\draw  (7,0) -- (7,1);
\draw (3.5,0.5) node{$t$};
\draw (10.5,0.5) node{$t+1$};

\draw  (4.0,-0.2) -- (9,-0.2);
\draw  (4.0,-1.2) -- (9,-1.2);
\draw (5.5, -0.7) node{$\lambda^c n^c$};
\draw (8, -0.7) node{$(1-\lambda^c) n^c$};
\draw (4.0,-0.2) -- (4.0,-1.2);
\draw (9,-0.2) -- (9,-1.2);

\draw  (5,-1.4) -- (10,-1.4);
\draw  (5,-2.0) -- (10,-2.0);
\draw (6, -1.7) node{$\lambda^f n^f$};
\draw (8.5, -1.7) node{$(1-\lambda^f) n^f$};
\draw  (5,-1.4) -- (5,-2.0);
\draw  (10,-1.4) -- (10,-2.0);

\draw[->] (11,0) -- (11,-2);
\draw (11,-2.5) node{Observe $s^c$, $s^f$,};
\draw (11,-3) node{order inputs: $n^c, n^f$};

\draw[->] (13,0) -- (13,-3.5);
\draw (13,-3.9) node{Shocks are realized: $\nu, \lambda^c, \lambda^f$};

\end{tikzpicture}
\label{fig:timing}
\end{figure}

Last, Figure \ref{fig:timing} shows the timing constraint firms face. Firms must decide how much of each of the foreign inputs to order before the demand and delivery time shocks are realized for the period. Firms place their orders according to their initial level of inventories and expected shocks. After the firm-specific demand and delivery time shocks are realized, $\{ \nu, \lambda^c, \lambda^f \}$, the fraction, $\lambda^c, \lambda^f $, of the orders arrive and can be used for production. Then, firms decide on the amount of each of the inputs, $x^d, x^c, x^f$, which they can use to produce and set the price, $p$. These choices, together with the law of motion of inventories, define the inventories for tomorrow. The rest of the order arrives at the beginning of the following period and is added to the inventory stock $s^{\prime i}$. 

The recursive problem for the final good producer is given by two Bellman equations, which correspond to the choices made within the timing constraint. For clarity, I drop the subscript denoting the specific firm in the unit continuum, $j$. The value function, $V(s^c, \: s^f)$, defines the optimal order of inputs, given the initial inventories of each input. Then, after shocks are realized and given the order of inputs, firms decide on the amount of inputs and labor to use in production, and set prices. This problem is described by the value function $\tilde{V}(s^c, \: s^f \:, n^c, \: n^f, \: \eta)$, where firms maximize present and future profits subject to 6 constraints: the demand from the consumer (equation \ref{demand}), the production function (equation \ref{tech}), the two constraints for the two foreign inputs (equation \ref{inputs}), and the law of motion for the close and far input inventories (equation \ref{lominv}).

\begin{equation*} 
\begin{split}
\label{value1}
 V (s^c, \: s^f) & = \: \:  \max_{ \{n^c, n^f \} }  \: E_{\eta} \Big[ \tilde{V}(s^c, \: s^f, \: n^c, \: n^f, \: \eta) \Big] \: \: \: \: \: \: \: \text{ where $\eta = (\nu, \: \lambda^c, \: \lambda^f)$} \\
  \tilde{V}(s^c, \: s^f \:, n^c, \: n^f, \: \eta) & =   \max_{ \{p, x^d, x^c, x^f, s^{\prime c}, s^{\prime f} \} }  \: \: p \: y(p) -  p^d \: x^d -  p^c \: n^c -  \: p^f \: n^f  + \beta \:  V (s^{\prime c}, \: s^{\prime f}) 
\end{split}
\end{equation*}   

Firms behave monopolistically by setting prices. If a firm is unconstrained in the amount of inventories it needs to meet the demand, then the price it sets is equal to a markup over the cost-minimizing marginal cost. If the firm is constrained by the available input quantity, it distorts its optimal price by changing the input mix. The binding constraint forces the firm to a suboptimal pricing strategy. As a consequence of the risk of foregone profits, the interaction between positive delivery times and demand shocks creates incentives for firms to hold inventories. Since firms have to wait for their inputs to arrive, while their demand changes every period, they need to store some of these inputs as inventories to ensure they will be able to meet their demand. Additionally, firms hold inventories because delivery times are stochastic, and firms do not have certainty regarding when they will arrive. On the other hand, inventories are costly; they depreciate at a rate $\delta$ and firms face a discount rate $\beta$. The cost of holding inventories creates a trade-off between the relative price and delivery times across inputs.

 %%%%%
\textbf{Partial equilibrium.  } 
We assume there is a unit continuum of the input producers that produce each variety, $j$. We abstract from modeling the problem of the inputs firms, and take as given the input prices. This is equivalent to assuming a constant return to scale technology for the input firms, where $x_j^i = A^i  \ell_j^i$ and thus $p^i=  w^i/A^i $ for $i \in \{ d, c, f\}$. All firms along the continuum have access to the same technology and do not face any shocks, so there is a unique price for the unit continuum of the inputs, $p^d, p^c, p^f$.  

%We also take as given the price index, $P$, and the output of the domestic economy, $Y$, which are part of the demand the final good firm faces every period.

%%%%%%%%%%%%%%
\subsection{Discussion of the mechanism of the model}
%% add here:
% (1) change in price of an input and effect of inventories
\paragraph{Tradeoff between price and delivery times.}
When selecting a production input bundle, firms consider not only the relative prices of inputs but also their inventory costs. The amount of inventory needed for each input depends on their delivery time distribution. Sourcing inputs with long delivery times means the firm receives only a small portion of the order placed today. To consistently meet demand, firms must maintain larger inventories of these inputs. Furthermore, if an input faces volatile delivery times, where the share of the input order that arrives today is uncertain, firms hold additional inventory of the input. Hence, inputs that face longer and more volatile delivery times are optimally more inventory-intensive. 

\paragraph{Change in the price of an input.}
As the price of an input rises, under our CES technology assumption, firms substitute away from that input and towards the remaining two inputs. If the price of an inventory-intensive input increases, then the total inventory stock decreases. Firms turn to inputs with shorter or less volatile delivery times that require less inventory. With rising tariffs on inputs from China, U.S. firms are shifting away from these inputs and increasing their reliance on domestic and the rest of the world inputs, which offer shorter and more stable delivery times. Consequently, total inventory holdings decrease.

% (2) effect of an increase in the variance of the dist of delivery times
\paragraph{Change in the volatility of delivery times.}
Firms' incentives to store inventories are driven by not only the mean but also the variance of the distribution of delivery times. An increase in the standard deviation of delivery times, raises the value of holding an additional unit of inventories. Thus, inputs facing longer and more frequent delays have higher inventory costs. However, the effect of a change in the volatility of delivery times on inventories is a priori ambiguous. On the one hand, as the firm substitutes away from that input, it needs to hold less of those inputs as inventories. On the other hand, higher delivery times increase the amount of inventories per utilized input unit the firm needs to hold.

As a consequence, the net change in total inventories depends on how easily inputs can be substituted and whether firms are moving toward more or less inventory-intensive options. As the delay increases, inventory requirements for the affected input are influenced by two opposing forces: firms need to stock more per unit used, but as they substitute away from this input, they require less inventories.

%If the delay of an input increases, then the \textit{inventories of this input} respond to two forces: firms need to hold more inventory per unit used, but as they substitute away from this input, they require less inventories. The change in inventory levels for the delayed input will hinge on the elasticity of substitution and the extent of the delay increase. Further the impact on \textit{total inventories} depends on whether firms substitute towards more or less inventory-intensive inputs. 

%%%
We use the model to examine the rise in delivery time volatility for both inputs that carry inventories. Recent global events have intensified delays faced by these inputs. We have seen an increase in supply disruptions due to factors such as COVID-19, port congestion, shortages of essential components, repeated blockages of key ocean trade routes, and escalating political conflicts. 
% should I give a preview of results?
In the quantitative section, we show that the rise in delivery time volatility for the two foreign inputs increases total inventories.

%% file: draft/4_quantification.tex
\section{Quantification}

To measure and study the costs of the rise in delays in the aggregate economy, in this section we describe how we choose the parameters for our quantitative analysis. We start by calibrating the model to match moments of the U.S. manufacturing sector in 2018. Then, we use data on the actual tariff change, the decline in the imports from China, and the rise in input inventories to calibrate the trade war with China that started in 2018, and the rise in delivery delays for inputs.

%\subsection{Benchmark calibration}
\subsection{Estimating the delivery time distributions}

Delivery times for each of the foreign inputs in the model are determined by the stochastic parameter $\lambda$. Equation (\ref{maxl}) shows the relationship between the delivery days observed in the data and $\lambda$. We assume $\lambda$ represents the proportion of days within a period the firm is able to use the order to produce. Thus, given the number of days in a period, T, $\lambda$ equals the proportion of days of the period the firm has the input in its warehouse and can use it to produce, $1 - $ delivery days/T.\footnote{If the delivery time is longer than the length of the period, then $\lambda$ is capped at a one-period delay, as common in the literature.} An alternative interpretation is that there is a continuum of firms ordering throughout the period. In which case, $\lambda$ represents the share of firms for which the order arrives before the period ends, and they are able to use the inputs to produce.
\begin{equation} 
\label{maxl}
\lambda_j^i = \max ( 0, \: 1 - \text{delivery days}_j^i /T ) \: \: \: \: \: \: \text{         for } i = {c , f} \text{ and firm } j  
\end{equation}  

The delivery days$_j^i$ for the inputs from China, $x^f$, and ROW, $x^c$, are drawn from two different distributions, $\text{days}_j^f \sim G^f(\mu_f, \sigma_f)$ and $\text{days}_j^c \sim G^c(\mu_c, \sigma_c)$. We assume each distribution is log-normal, and estimate the geometric mean to be equal to the mean of delivery days in the data. Further, we calculate the geometric standard deviation such that $2/3$ of the distribution lies within the observed delays. In this case, delays can be either early or late deliveries relative to the expected delivery date.

To estimate the delivery day distribution for the inputs from China, we use data on ocean transportation times. Around $80\%$ of total imports that arrive from China use ocean transportation, which takes on average 30 days to arrive to the U.S., according to the logistics company \textit{Freightos}. The expected delays reported by \textit{Sea Intelligence}, defined as days before or after the expected delivery, are approximately $10$ days. Thus, the geometric mean, $\mu_f$, equals 30, and the standard deviation equals $\sigma_f = \frac{\mu_f + 10}{\mu_f} = 4/3$.

To estimate the parameters of the distribution of delivery days for the ROW imports, we compare the distance traveled by imports from China to the import weighted distance of imports from ROW. According to the population-weighted distance reported by CEPII, imports from China travel $11,613$ km. We measure the import weighted distance traveled by ROW in 2018, $D = \sum_i \frac{D_i M_{i, 2018}}{\sum_j M_{j, 2018}}$, where $D_i$ is the distance traveled by imports $M_i$ from country $i$, and we sum over all imports except China. The distance traveled by ROW is equal to 0.65 of the kilometers traveled by imports from China in 2018. Then, we define the average delivery days of ROW to be proportional to the days from China, and the geometric mean is equal to $\mu_c = (30)(0.65) = 19.5$. Similarly, the delays would be equal to a fraction $0.65$ of the 10 day delay observed in the trade route between the U.S. and China. Thus, the geometric standard deviation is the same across foreign inputs, and equals $\sigma_c = \frac{\mu_c + 6.5}{\mu_c} = 4/3$.

\subsection{Assigned parameters}
 Panel C in Table \ref{table:t1} summarizes the parameters that we take from the literature. A period is equal to a quarter, T = 90 days. The discount factor is set to $\beta = 0.96^{1/4}$, which corresponds to a $4\%$ annual interest rate. The depreciation rate for inventories, $\delta$, equals $7.5\%$, which implies a $30\%$ annual rate following the work by \citet{richardson}. In line with the literature, we set the elasticity of demand for a firm's variety of the final good, $\epsilon$, equal to 4 to generate markups of $1.33$.
%The parameter values summarized in Table \ref{table:t1} are calibrated to match moments of the U.S. manufacturing in 2018.
\subsection{Jointly determined parameters}
The remaining parameters are jointly determined to match four moments from the data. First, from the firm's technology in equation (\ref{tech}), we set the weight of inputs from China, $\theta_f$, and the weight of inputs from ROW, $\theta_c$, to match the imports over sales from each input observed in the data in 2018. Here, $\theta_f = 0.089$ and $\theta_c = 0.444$. Second, we assume the idiosyncratic per period demand shocks the firms faces, $\nu$, are drawn from an independent and identically distributed log-normal distribution, where the standard deviation, $\sigma_{\nu}$, equals 0.638 to match the input inventories over quarterly sales in 2018. We define input inventories as the sum of materials and supplies plus the work-in-process inventories as defined by the U.S. Census Bureau. 

Last, we calibrate the elasticity of substitution between inputs, $\sigma$, to match the decline in imports from China over sales from 2018 to 2024. To do so, we first measure the import-weighted rise in tariffs for imports from China, as in \citet{amiti}. Using tariff data from the U.S. International Trade Commission (USITC) and the U.S. Trade Representative (USTR), and trade volumes from the U.S. Census Bureau, the initial average weighted tariffs in 2018 were $3\%$, and had increased to $18\%$ in 2024, shown in Table \ref{table:tariffs}. Given the rise of $15$ percentage points in the net price of imports from China, we then calibrate the elasticity of substitution required to match the $3.23$ percentage point decline in imports from China over sales observed from 2018 to 2024. The elasticity, $\sigma$, equals 4.51.

\begin{center}
\resizebox{410pt}{!}{
\begin{threeparttable}
\caption{Parameters}
\begin{tabular}{ l c c | c c c  }
\multicolumn{6}{l}{Quarterly model, $T = 90$}	\\
\multicolumn{6}{l}{ } 		\\
\multicolumn{6}{l}{\textbf{Panel A. Calibrated parameters}	}	\\
Parameter					& 					& 	Value 	&  Moment			& Model	& Data 	 \\
\hline 
\hline														
Weight inputs China  		& $\theta^f$		& $0.089$		& inputs China / sales in 2018 		& $8.89\%$ 	& $8.89\%$  		\\
Weight inputs ROW 			& $\theta^c$		& $0.444$		& inputs ROW / sales in 2018 		& $27.75\%$ 	& $27.74\%$  		\\
Elasticity inputs  			& $\sigma$		& $4.511$		& $\Delta_{18-23}$ inputs China / sales	& $-3.24\%$ 	& $-3.23\%$  		\\
Variance of demand 			& $\sigma_{\nu}$	& $0.638$		&input inventory / sales in 2018	& $28.86\%$ 	& 	$28.86\%$  	\\
            			& \multicolumn{5}{l}{$\nu_j \sim \mathcal{LN}(0, \sigma_{\nu})$   } \\

\multicolumn{6}{l}{} \\
\multicolumn{6}{l}{\textbf{Panel B. Estimated parameters }} 		\\
Parameter				& 		 	&  \multicolumn{4}{c}{Value}			 \\
\hline
\hline
Delivery times China		&$\lambda_f$	& \multicolumn{4}{c}{$ \text{days}^f \sim \mathcal{LN}(30, 1.33)$ }	 \\
Delivery times ROW		& $\lambda_c$	& \multicolumn{4}{c}{$ \text{days}^c \sim \mathcal{LN}(19.5, 1.33)$ }  	\\	
					&			& \multicolumn{4}{c}{$\lambda = \max( \: 0, \: 1 - \text{days}/T) $ }	\\
Tariffs on China	&	$\Delta \tau$		& \multicolumn{4}{c}{ $\tau_{17} = 3.0\%$ and  $\tau_{24} = 18.1\%$, then $\Delta_{17-24} \: \tau = 15.1\% $ } \\
\multicolumn{6}{l}{} \\
\multicolumn{6}{l}{\textbf{Panel C. Predetermined parameters }} 		\\
Parameter					& 					& 	Value 	&  \multicolumn{3}{l}{Comment}			 \\
\hline
\hline
Elasticity of sub. $y$		&$\epsilon$	& $4.0$			& \multicolumn{3}{l}{Markups = $4/3$ }	 \\
Interest rate		& $\beta$		& $0.96^{1/4}$	& \multicolumn{3}{l}{$4\%$ annual interest rate}  	\\	
Depreciation rate		& $\delta$		& $0.075$			& \multicolumn{3}{l}{$30\%$ annual rate (inventory holding costs)} \\
\end{tabular}
\label{table:t1}
     \end{threeparttable}
     }
\end{center}
%maybe in results?
%\subsection{Estimating the trade war and the rise in delays}

%% file: draft/4_results.tex
\section{Quantitative results}
In this section, we describe how we quantify the costs of the rise in volatility in terms of the rise in delays for inputs. Further, we measure the cost of delays in terms of the change in output, prices, and sourcing choices. To do so, first, we calibrate the rise in delivery delays to match the observed rise in input inventories over sales. Then, to distinguish the cost of delays from the cost of the rise in tariffs, we compare the stationary distribution for different counterfactuals. Last, we show the tariff equivalent measure of a 10-day rise in delays.

\subsection{Quantifying the rise in delays}
To quantify the rise in delays, we internally calibrate the change in the relative price of inputs from China and the standard deviation of the delivery times for foreign inputs from China and ROW. We do so by matching the decline in the manufacturing imports from China over sales and the rise in input inventories over quarterly sales observed in the data. We start by computing the steady state in 2024. To do so, we rely on the initial calibration strategy where the elasticity of substitution, $\sigma$, is calibrated to match the decline in imports from China over sales observed in 2024, given the rise in tariffs of $15$ percentage points. Then, we calibrate the change in the delays, $\Delta \text{delays}$, to match the input inventory data in 2024. To do so, we adjust the geometric standard deviation of the delivery days of both the inputs from China and ROW as follows. We find that a rise in delays of an additional 21.25 days matches the 2024 inventory data.
  \begin{equation}
\label{delays}
\begin{split}
\sigma^c = \: & \: \frac{\mu^c + 6.5 + \Delta \text{delays}}{\mu^c} \\
\sigma^f = \: & \: \frac{\mu^f +10.0 + \Delta \text{delays}}{\mu^f} \\
\end{split}
\end{equation}

\begin{figure}[!ht]
\centering
\caption{Rise in delays}
\begin{subfigure}{0.5\textwidth}
\centering
\includegraphics[width=6.5cm]{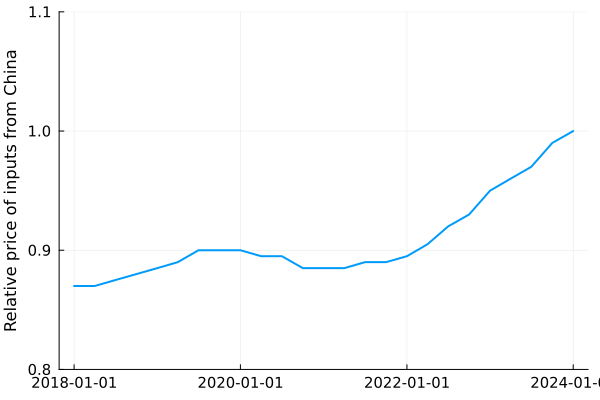}
\caption{\footnotesize{Rise in tariffs for imports from China}}
\end{subfigure}\hfill
\begin{subfigure}{0.5\textwidth}
\centering
\includegraphics[width=6.5cm]{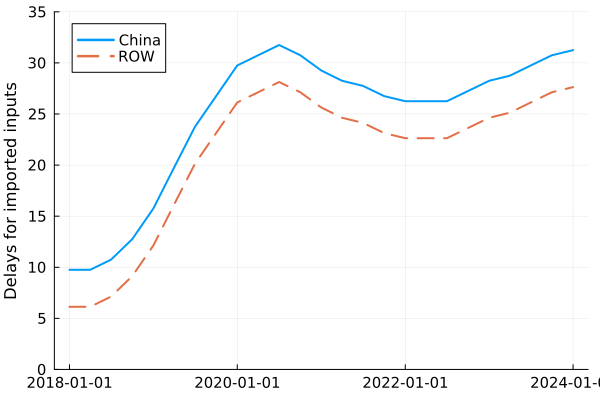}
\caption{\footnotesize{Rise in delivery delays for imported inputs}}
\end{subfigure}
\medskip
\begin{subfigure}{0.5\textwidth}
\centering
\includegraphics[width=6.5cm]{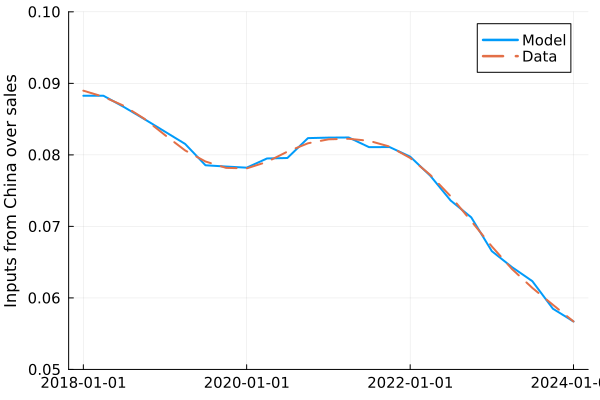}
\caption{\footnotesize{Match inputs from China over sales}}
\end{subfigure}\hfill
\begin{subfigure}{0.5\textwidth}
\centering
\includegraphics[width=6.5cm]{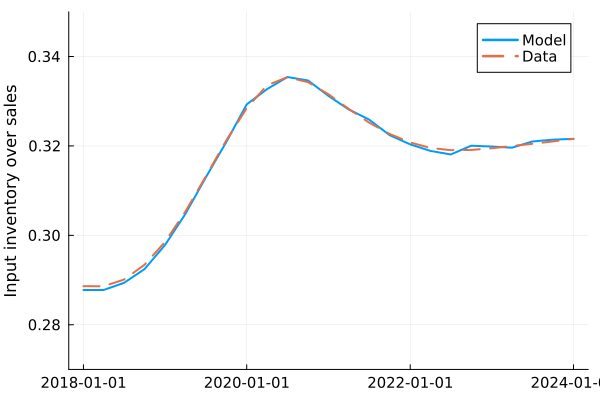}
\caption{\footnotesize{Match the increase in input inventory over sales}}
\end{subfigure}
\caption*{\footnotesize{}}
\label{fig:calib1}
\end{figure}

Next, we obtain the calibrated change in delays, $\Delta \text{delays}_t$, and the price of the input from China, $\tau p^f_t$, for each quarter from 2018 to 2024. They are calibrated to match HP filtered trend of the manufacturing imports from China over sales and the inputs inventories over sales in every period, as shown in panels (c) and (d) in Figure \ref{fig:calib1}. To do so, we use backward induction under the assumption of perfect foresight along the transition. Panel (a) shows the rise in the price of inputs from China for each quarter, and panel (b) shows the rise in delivery delays. The rise in delays has an initial peak of 21.4 days in the second quarter of 2020, following the rise in the inventory data. Then, delays decline to around an additional 16 days during the first quarter in 2022, but they increase again to $21.2$ days in 2024. Delays remain high in 2024 because as imports from China decline, delays need to continue to rise to match the rise in inventories.

\subsection{Comparing the model delivery days to the data.}
\begin{figure}
\centering
\caption{Comparing our measure on delivery days to ISM}
\begin{subfigure}{.475\textwidth}
  \centering
  \includegraphics[width=\textwidth]{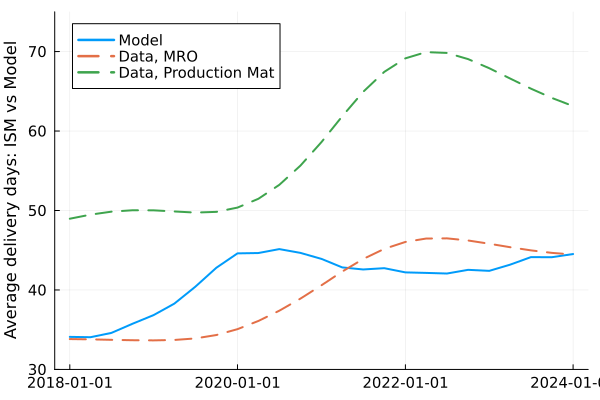}
  \caption{Model, MRO, and Production Material }
\end{subfigure}%
\begin{subfigure}{.475\textwidth}
  \centering
  \includegraphics[width=\textwidth]{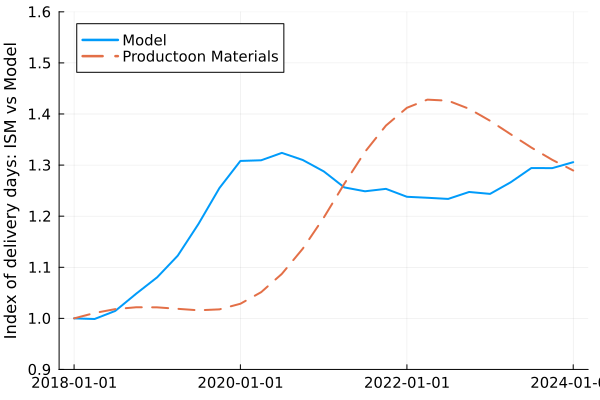}
  \caption{Growth of model and production materials}
\end{subfigure}
\label{fig:ismmodel}
\end{figure}

We compare our model measure of delivery times to the data on lead times reported by the ISM. To do so, we replicate the ISM survey procedure in our model.  The ISM lead times are computed by measuring the number of firms that report their lead time in different bins of time, for 5, 30, 60, 90, 180, or 365 days or more. We take a large number of draws of each of the domestic and foreign distributions and compute the mean days for each trade flow. The mean of the days follows the ISM methodology, where we categorize each draw in a bin, and then take the average across bins. Then, we aggregate them using the share of inputs from China and ROW over total imports. 

\begin{equation} 
\label{daysism}
 \text{days}^{\text{model}}_t = \frac{\text{imports}^c_t}{\text{total imports}_t} \: \frac{ \sum_j \text{bin}_j \sum_{i \in \text{bin}_j}^N \text{days}^c_i }{N} \: + \: \frac{\text{imports}^f_t}{\text{total imports}_t}   \: \frac{  \sum_j \text{bin}_j \sum_{i \in \text{bin}_j}^N \text{days}^f_i }{N} 
\end{equation}  

where $\text{days}^c_i$ and $\text{days}^f_i$ are draws from the estimated log-normal delivery time distributions, and $\text{bin}_j$ is the days in a given bin defined by ISM. Figure \ref{fig:ismmodel} shows the model comparison to the two ISM series for maintenance, repair, and operating supplies (MRO), and production materials. Panel (a) shows the model delivery days computed using Equation (\ref{daysism}) and the lead times reported for MRO supplies and Production Materials. The model-based initial delivery times for foreign imports equal 34 days, which is similar to the days reported for MRO supplies in 2018. The model-based days grow faster than the ISM days at the beginning, but the total growth from 2018 to 2024 equals 11 days for both the model and MRO days.\footnote{A simple explanation for the different dynamics in the model and the data is that we do not include any aggregate change in demand. As a consequence, the increase in inventories partly driven by changes in aggregate demand during the Covid pandemic is loaded on delays in the model. As changes in demand subside, the model correctly captures the rise in delays that explains the new level of inventory-to-sales ratio.} The lead times for Production Materials start at 49 days in 2018 and grow by 14 days. While the level is off by 7 days with respect to the Production Materials, the growth in both series equals 13 days and 11 days, respectively. Panel (b) shows the growth of the Production Materials lead times and our model-based measure of days. 
 
%(ii) and with  “Input Sourcing Under Supply Chain Risk: Evidence from US Manufacturing Firms” (with Esposito and Heise)  -- ANOTHER  ONE?
%(iii) next section: Results compare to: The aggregate effects of global and local supply chain disruptions: 2020–2022

\subsection{Cost of the rise in tariffs and delays}
In our model, the combination of the rise in a 21-day delay and an increase in tariffs for inputs from China of $15$pp leads to a drop in output of $7.3\%$ and an increase in prices of $1.8\%$. Inventories are influenced by two opposing factors. On one hand, the increase in tariffs prompts firms to shift away from Chinese inputs, which have the longest delivery times, reducing the need for inventories. On the other hand, as delays grow, firms are compelled to hold more inventories per unit used. Given our calibration strategy, where we match the rise in inventories, the impact of delays will outweigh the substitution effect, resulting in an $11\%$ increase in total inventories relative to sales. Isolating the effects of delays, we show that output drops by $2.6\%$ and the final good prices increase by $0.4\%$. In this case, inventories rise $17\%$. 

To disentangle the cost of the rise in tariffs from the cost of the rise in delays, we compare the moments of the stationary distributions of different economies, shown in Table \ref{table:levelt}. The first column shows the benchmark economy, obtained by matching the moments of U.S. manufacturing in 2018. Here, we compute the equilibrium using the low tariffs observed in 2018 and no change in delays. The second column isolates the role of the trade war. We consider the steady state, where delays remain at the initial level, but firms face the increase in tariffs of $15$pp (\textit{tariffs}). The third column outlines the moments of an economy in which delays increase by $21$ days, but tariffs remain at their 2018 level (\textit{delays}). The last column shows the effect of both the tariff increase and the rise in delays that match the moments of the 2024 economy. The last three columns of Table \ref{table:levelt} show the comparison across different economies. Column 5 shows the impact of the increase in tariffs through the change between columns 2 and 1. Column 6 quantifies the cost of delays by showing the change between columns 3 and 1. Last, column 7 shows the change in the U.S. economy between 2024 and 2018. Appendix \ref{appendix:rob} shows the same exercise computed for the average increase in tariffs of $19.7$ percentage point increase (see Table \ref{table:tariffs}), where results remain similar.
We discuss each counterfactual in turn. 

\textbf{Cost of the rise in tariffs.}
Column 5 of Table \ref{table:levelt} isolates the cost of the rise in tariffs, where output declines $5.1\%$ and the final good prices increase $1.4\%$. As the tariffs imposed in imports from China increase, firms substitute away from these inputs and towards ROW and domestic inputs, which grow $5.2\%$ and $4.9\%$ respectively. The higher cost of input bundle raises the final price, and output declines in response. As firms use less of the input from China, the inventories for these inputs decrease by $35\%$, yet as firms increase the use of ROW inputs, they increase their stock of this type of inventories by $5.2\%$. As firms substitute towards domestic and ROW inputs, which are more readily available and require fewer inventories per unit used, the total inventories over sales drop $5.7\%$. As expected, when firms reduce their dependence on imports from far away sources, inventories decrease in response. 

\textbf{Cost of the rise in delays.}
A 21-day delay leads to an output decline of $2.6\%$, and the final good price increases $0.4\%$, as shown in column 6. As the standard deviation of the delivery times increases, the added uncertainty in the share of the order they can access before production increases the firm's incentives to hold additional inventories per unit of input used. Inventories are costly, which increases the price. Further, the rise in volatility increases the share of firms that are constrained at a given period, which forces the firm to suboptimally increase their reliance on the more expensive domestic inputs, which also raises their price and lowers output. These results are comparable to \citet{akk2021} who report a $3\%$ drop in output and a $2\%$ increase in final prices due to delays. As the delays for foreign inputs increases, firms substitute away from inputs from China and ROW ($-1.9\%$ and $-2.1\%$ respectively), and towards domestic inputs. Even though firms are using less of the foreign inputs, inventories for each of these inputs increase, as the delays raise the need for inventories per unit used. Thus, total inventories increase by $17.1\%$.

\textbf{Cost of both the rise in tariffs and delays.} The output costs considering both the increase in tariffs and delays equals $7.3\%$, and the final good prices rise $1.8\%$ from 2018 to 2024. In this case, the combined effect of increased tariffs and delays drives up the final price, causing a sharper drop in output. Inputs from China observe a steep decline because of the direct increase in their price and their more volatile delivery times. Yet, their inventory holdings face opposing forces. As firms substitute away from these inputs, they carry fewer inventories, but the increase in delays raises the need for inventories per unit used. Quantitatively, the substitution effect is stronger, and inventories of inputs from China decline by $23\%$. Inputs from ROW increase by $3\%$, since the effect of firms substituting towards these inputs is larger than the drop of the inputs due to the rise in delays. Thus, inventories over sales of ROW inputs increase by $24\%$, since firms use more of these inputs and have to carry more inventories per unit used. The increase in ROW inventories mitigates the fall of the inventories from inputs from China, and total inventories increase by $11\%$.

\begin{table}
\resizebox{480pt}{!}{
\begin{threeparttable}
 \captionof{table}{Quantify costs of tariffs and delays}
\begin{tabular}{l | c c c c c c c }
    & (1) & (2) & (3) & (4) & (5) & (6) & (7) \\
			& \textbf{2018}			& \textbf{Tariffs}			& \textbf{Delays}		& \textbf{2024}		& \textbf{Tariffs vs}			& \textbf{Delays vs}		& \textbf{2024 vs}		 \\
                     & \textbf{Benchmark} &  \textbf{$\uparrow \tau^c$} & \textbf{$\uparrow$ delay} &  \textbf{$\uparrow \tau^c$ + delay} & \textbf{2018} & \textbf{2018} & \textbf{2018} \\
\hline
\hline
        Tariff             & $\tau^c p^c = 0.87$ 	& $\tau^c p^c = 1.0$ 	& $\tau^c p^c = 0.87$ & $\tau^c p^c = 1.0$ & & &  \\
         Change delays         &  $\Delta$ delay = 0	& $\Delta$ delay = 0	& $\Delta$ delay = 21 & $\Delta$ delay = 21  & & & \\
        Output                   & 0.915 & 0.868 & 0.891 & 0.848 & -5.1\% & -2.6\% & -7.3\% \\ 
        Prices                   & 1.350 & 1.368 & 1.355 & 1.374 & 1.4\% & 0.4\% & 1.8\% \\ 
        Inputs China/sales         & 0.089 & 0.057 & 0.087 & 0.057 & -35.4\% & -1.9\% & -36.1\% \\ 
        Inventories China/sales & 0.078 & 0.051 & 0.090 & 0.060 & -34.9\% & 14.6\% & -23.2\% \\ 
        Inputs ROW/sales         & 0.278 & 0.292 & 0.272 & 0.286 & 5.2\% & -2.1\% & 3.0\% \\ 
        Inventories ROW/sales    & 0.21 & 0.221 & 0.248 & 0.261 & 5.2\% & 18.0\% & 24.0\% \\ 
        Inputs domestic/sales   & 0.368 & 0.386 & 0.376 & 0.393 & 4.9\% & 2.0\% & 6.7\% \\ 
        Inventories/sales   & 0.289 & 0.272 & 0.338 & 0.321 & -5.7\% & 17.1\% & 11.2\% \\ 
    \end{tabular}
\label{table:levelt}
\begin{tablenotes}
\item \footnotesize{Columns 1 to 4 in the table report the average of the stationary distribution for different variables. Column one represents the benchmark economy of 2018 with the low initial tariffs and no additional delays. Column two isolates the effects of tariffs, and computes the steady state average of an economy with the raise in tariffs and low delays. Column three shows an economy with high delays and the initially low tariffs. Then, column four shows an economy with both the rise in tariffs and delays. The last three columns compares the economies relatively to the benchmark 2018 economy to show the effects of tariffs, in column 5, the effect of delays, column 6, and both changes in column 7.}
\end{tablenotes}
     \end{threeparttable}
     }
\end{table}

%% file: draft/5_conclusions.tex
\RaggedRight\section{Conclusions}

In this paper, we study the effect of rising delays in supply chains and their economic cost. We use a model of international sourcing matched to import flows and the stock of inventories over time. Accounting for the U.S.-China trade war that started in 2018, we estimate a rise of 21 days in the average U.S. import shipping delay. The combination of rising tariffs and delays induces an output loss of 7.3\% and a price increase of 1.8\%. We decompose these into the part driven by the trade war (5.1\% and 1.4\%, respectively) and the contribution of rising delivery delays (2.6\% and 0.4\%). 

\pagebreak

%% file: draft/6_appendix.tex
\begin{comment}
\raggedright \section{Tariff Change}
\label{tariff}
We compute the average tariff change between the US and China using tariff data from the USITC and USTR, together with trade volumes from the US Census Bureau. 

Formally, we define the weighted tariff change as 
\begin{align}
    \Delta \tau_t =\frac{\sum_i \Delta \tau_{it} \bar M_{i}}{\sum_i \bar M_i}, 
\end{align}
with $\Delta \tau_{it}$ is the change in tariff applied to product $i$ between 2017 and 2019, and $\bar M_i$ is the US import value of Chinese product $i$ in January 2017. 

We obtain a weighted change in tariff of 15\%, which we use in our quantitative model.

\begin{table}[h!]
\centering
\begin{tabular}{@{}lcc@{}}
\toprule
\textbf{Tariff Rate} & \textbf{Weighted Average} & \textbf{Average} \\ \midrule
2017m1 & 3\% & 5\% \\
%2019m1 & \textbf{0.09203777} & 0.1341093 \\
%$\Delta$ (2017m1-2019m1) & \textbf{0.06206906} & 0.085879 \\
2019m10 & 18\% & 25\% \\
$\Delta$ (2017m1-2019m10) & {15\%} &20\% \\ \bottomrule
\end{tabular}
\caption{Tariff rate changes over time.}
\end{table}

\end{comment}

\raggedright 
\section{Empirical Analysis}
\label{appendix:data}
Here, we provide additional evidence for the two empirical observations motivating the paper. First, we show that the decline in manufacturing imports from China and the rise in manufacturing inventories over sales from 2018 to 2024 is present across the NAICS three digit sectors. Then, we include information on the transportation costs for air and ocean shipping that are used in creating the GSCPI. Here, we show the rise in transportation costs that started in 2020, which remains high for 2024. Last, we show the increased observed in the average lead times for capital expenditures, maintenance, repair, and operating supplies, and for production materials. 

\begin{figure}
\centering
\caption{Decline in imports from China across sectors}
\begin{subfigure}{.475\textwidth}
  \centering
  \includegraphics[width=\textwidth]{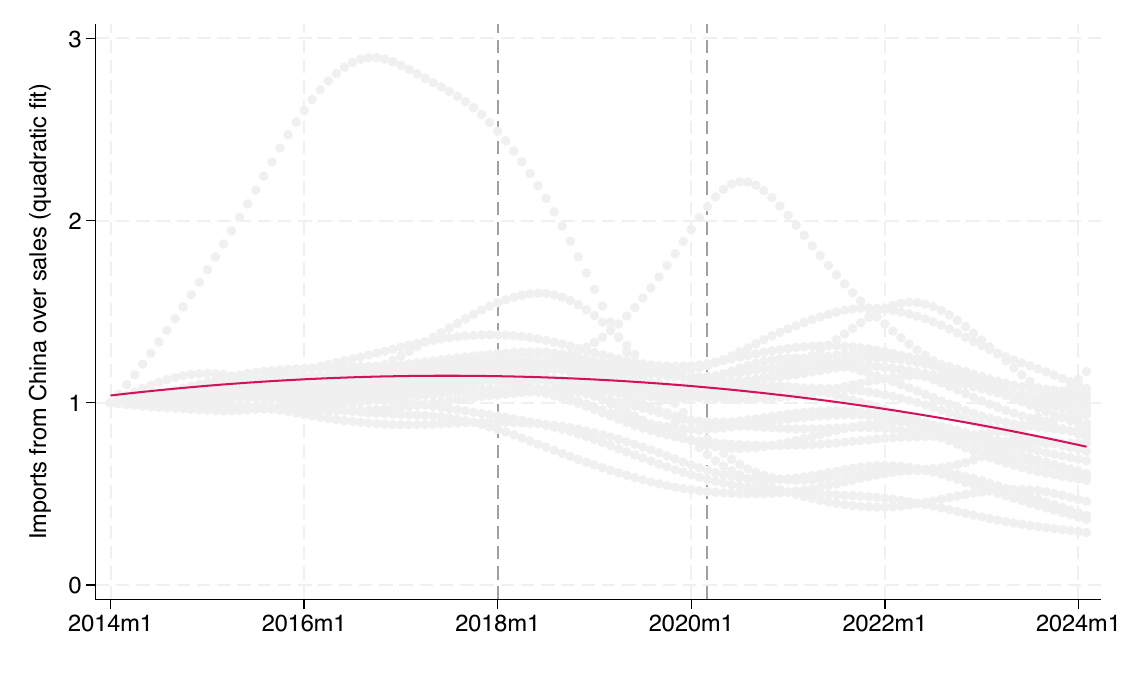}
  \caption{Decline in imports from China}
\end{subfigure}%
\begin{subfigure}{.475\textwidth}
  \centering
  \includegraphics[width=\textwidth]{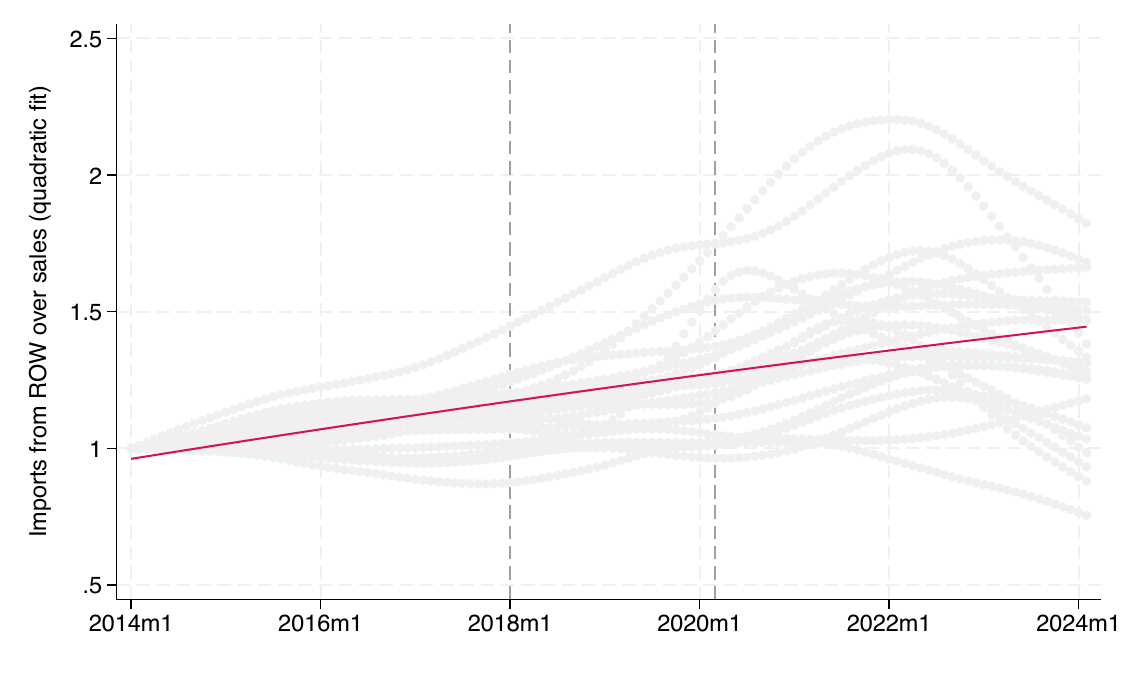}
  \caption{Rise in imports from ROW}
\end{subfigure}
\label{fig:imports2}
\caption*{The figures show the HP filtered trends for both the imports from China over sales and the imports from ROW over sales for the three digit NAICS industries. The solid line show the quadratic fit across the industries.}
\end{figure}

\begin{figure}
\centering
\caption{Decline in total imports over sales across sectors}
\begin{subfigure}{.475\textwidth}
  \centering
  \includegraphics[width=\textwidth]{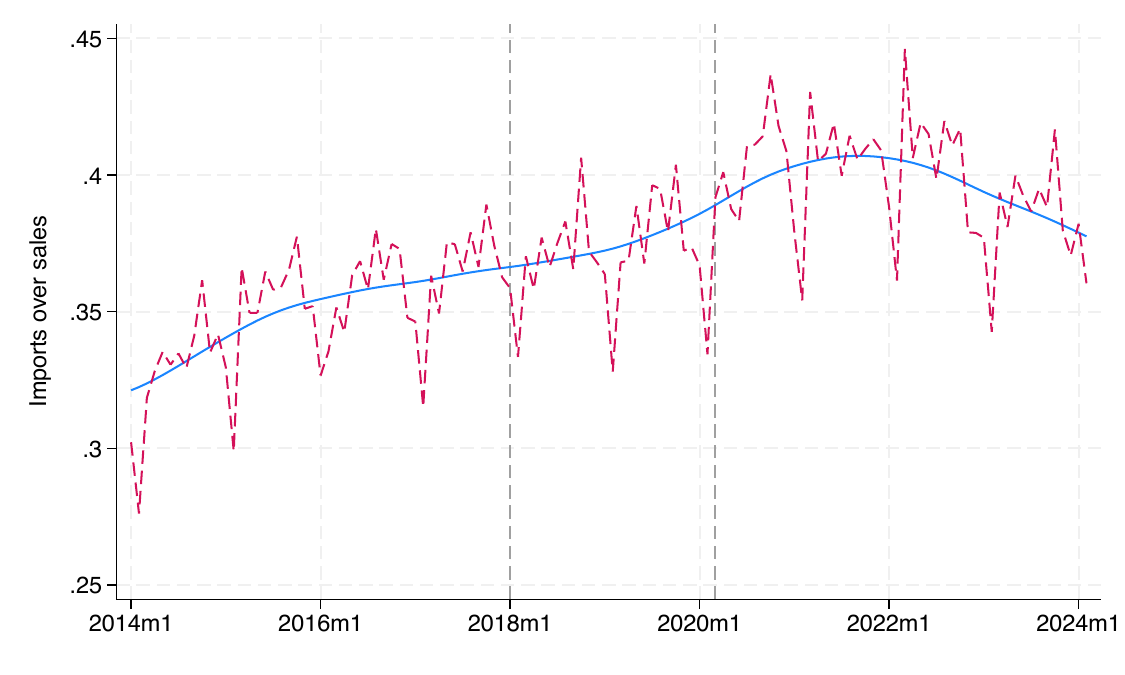}
  \caption{Decline aggregate total imports over sales}
\end{subfigure}%
\begin{subfigure}{.475\textwidth}
  \centering
  \includegraphics[width=\textwidth]{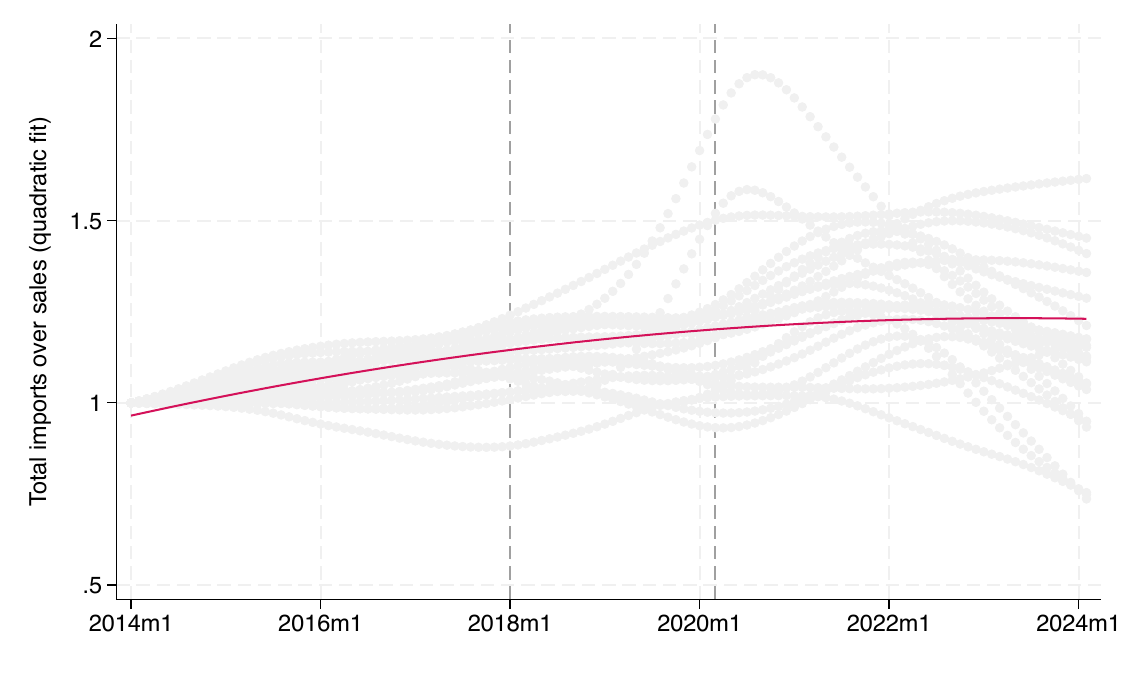}
  \caption{Total imports over sales across sectors}
\end{subfigure}
\label{fig:imports3}
\caption*{Panel (a) shows the total monthly manufacturing imports over sales and the HP filtered trend, which shows a decline starting in 2022. The decline is observed across sectors, a shown in Panel (b). The figure plots the HP filtered trend for the total imports over sales for the three digit NAICS industries and the solid line show the quadratic fit across the industries.}
\end{figure}

\begin{figure}
\centering
\caption{Rise in input inventories over sales across sectors}
  \centering
  \includegraphics[width=0.5\textwidth]{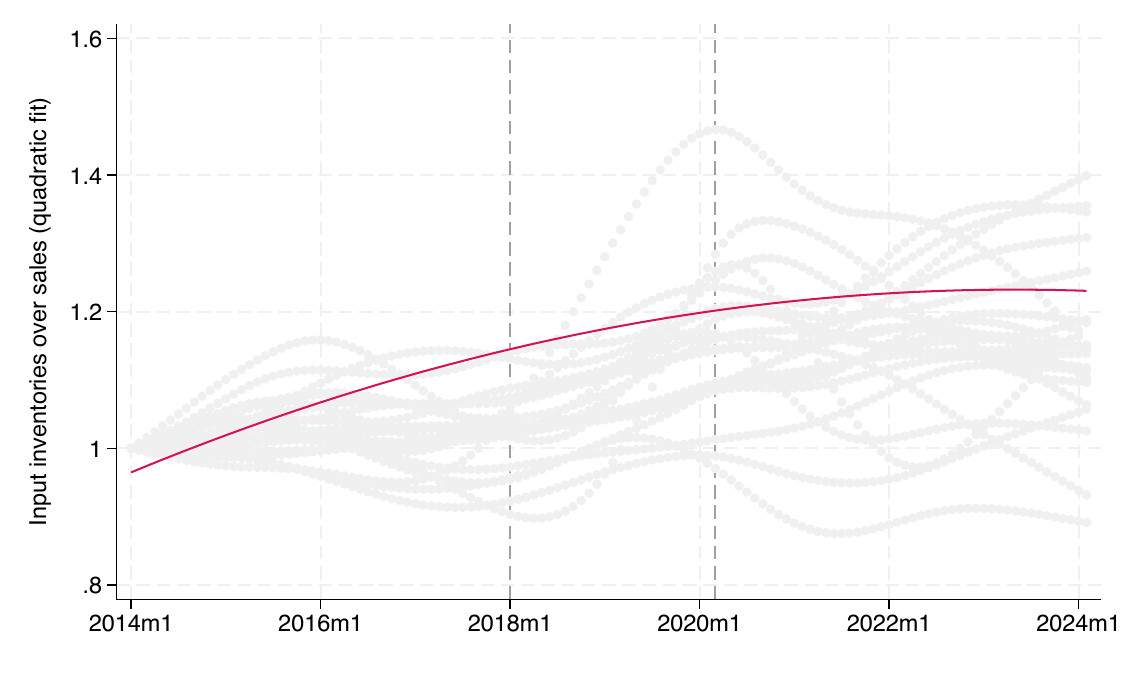}
  \caption{Decline in imports from China}
\label{fig:inv2_app}
\caption*{The figures show the HP filtered trend for the input inventories over sales across sectors and the quadratic fit, which shows the rise across sectors.}
\end{figure}

In Figure \ref{fig:imports2} we show the decline in manufacturing imports from China and the rise in ROW imports over sales across NAICS three digit sectors. The figures plot the HP filtered trends for each sector in gray, and the solid lines represent the quadratic fit for each series. Further, Figure \ref{fig:imports3} shows the decline in the total manufacturing imports over sales that starts in 2022. Total imports decline due to the sharp decrease in imports from China, which is not compensated by the growth in the ROW manufacturing imports after 2022. This trend is observed across manufacturing sectors, as shown in Panel (b) in Figure \ref{fig:imports3}. Next, Figure \ref{fig:inv2_app} shows the rise in manufacturing input inventories over sales across the NAICS three digits sectors.

\begin{figure}
\centering
\caption{Increase in transportation costs}
\begin{subfigure}{.475\textwidth}
  \centering
  \includegraphics[width=\textwidth]{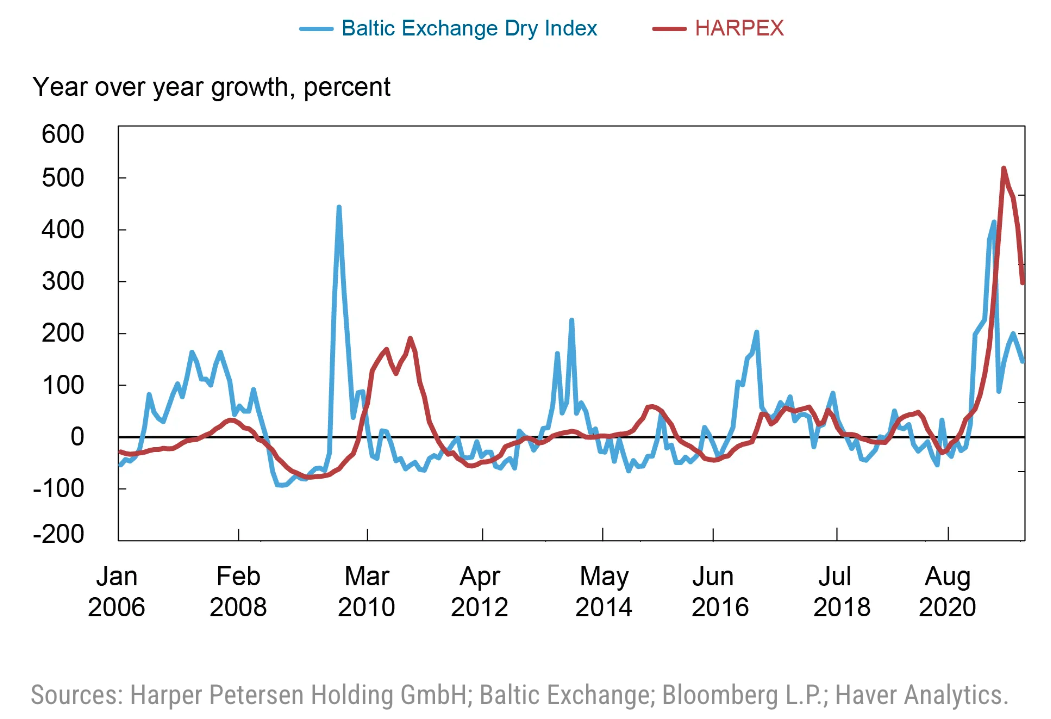}
  \caption{Increase in ocean shipping rates}
\end{subfigure}%
\begin{subfigure}{.475\textwidth}
  \centering
  \includegraphics[width=\textwidth]{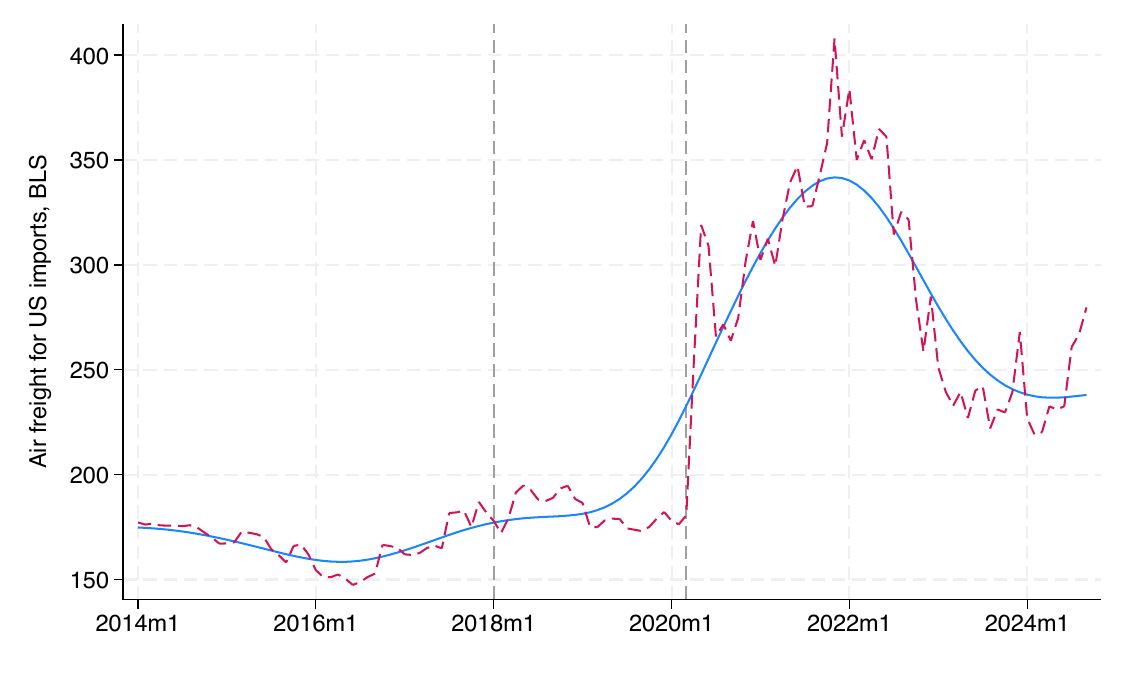}
  \caption{Increase in air freight rates}
\end{subfigure}
\label{fig:transport}
\caption*{Panel (a) shows the Baltic Exchange Dry Index and the Harpex Index, which have a sharp increase in 2020. Panel (b) shows the import air freight index, reported by the Bureau of Labor Statistics, and its the HP filtered trend. All three indices show the increase in 2020.}
\end{figure}

Figure \ref{fig:transport} shows the rise in transportation costs in 2020 for ocean and air used to construct the GSCPI. Panel (a) shows the Baltic Exchange Dry Index, which measures the cost of shipping raw materials through a composite of dry bulk time charter costs, and the Harpex Index, which measures worldwide price changes in the charter market for container ships. Both of these indices show the rise in ocean transportation freight rates during 2020. Panel (b) shows the rise in import air freight index reported by the Bureau of Labor Statistics, which measures the change sin air rates for air transportation of freight to the U.S. There is a sharp increase in air freight rates starts in 2020, and we see a decline in 2022, but by 2024 the level is far above the air freight cost rates of 2018.

\begin{figure}
\centering
\caption{Average lead times reported by the ISM}
\begin{subfigure}{.475\textwidth}
  \centering
  \includegraphics[width=\textwidth]{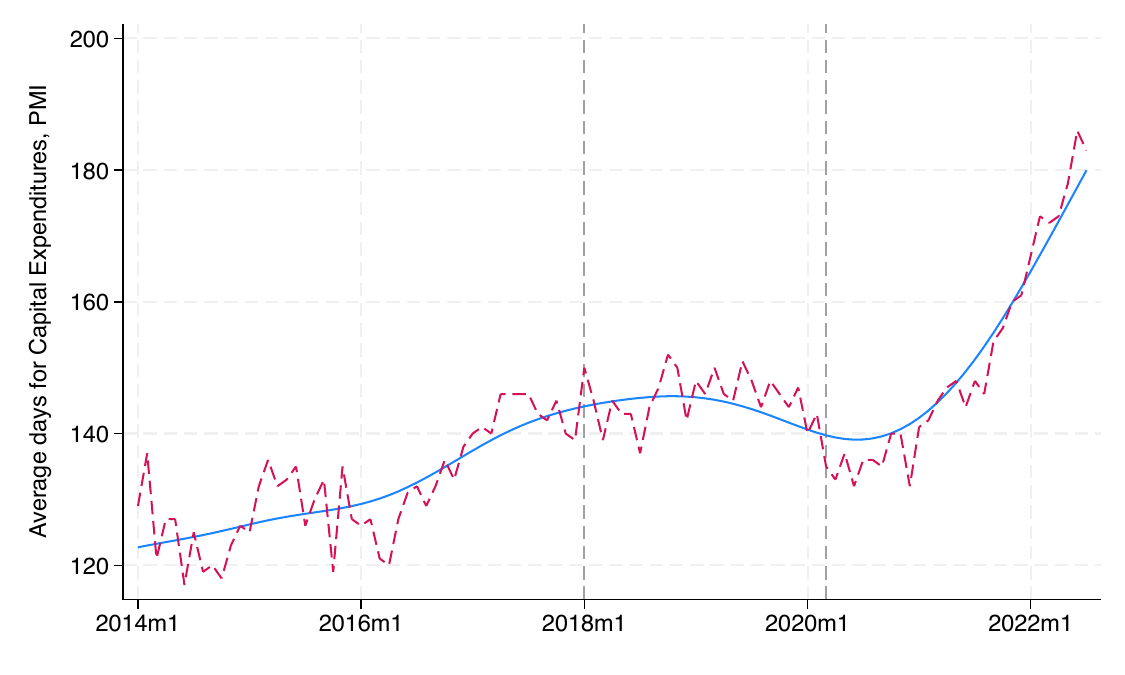}
  \caption{Average lead times in capital expenditures}
\end{subfigure}%
\begin{subfigure}{.475\textwidth}
  \centering
  \includegraphics[width=\textwidth]{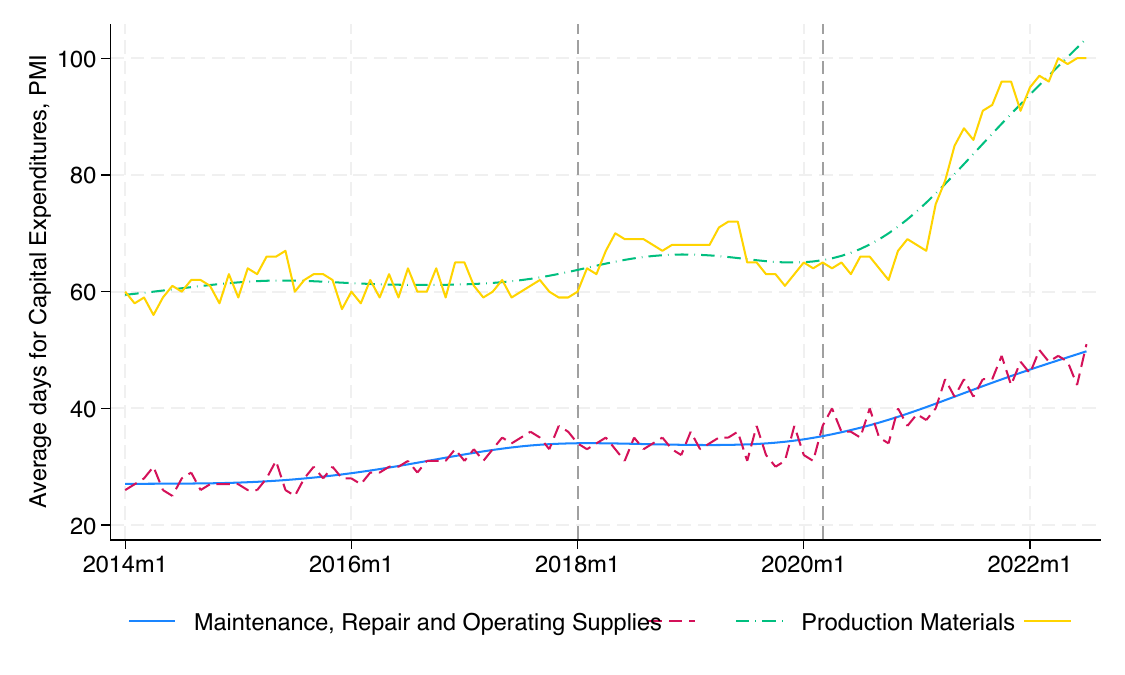}
  \caption{Production materials, and maintenance, repair and operating supplies}
\end{subfigure}
\label{fig:leadtimes}
\caption*{Panel (a) shows the average lead times for capital expenditures reported by the ISM, and panel (b) shows the times for production materials, and maintenance, repair and operating (MRO) supplies. All series show the rise in lead times in 2020. }
\end{figure}

Last, Figure \ref{fig:leadtimes} shows the average lead times reported by the ISM. The lead times are computed by measuring the number of firms that report their lead time in different bins of time: between 5, 30, 60, 90, 180, or 365 days or more. Lead times are reported for capital expenditures, production materials, and maintenance, repair and operating (MRO) supplies. Lead times across supplies increase sharply in 2020 and remains high for 2024.

%%%%%%%%%%%%%%%%%%%%%%%%%%%%%%%%%%%%%%%%%%%%%%%%%
\section{Robustness exercise}
\label{appendix:rob}
In this section, we include a robustness exercise to include a larger increase in tariffs, than the $15$ percentage points we use in the main text. Here, we use the average increase in tariffs reported in Table \ref{table:tariffs} of $19$ percentage point increase. The results hold across both exercises.

We first recalibrate the benchmark model to fit the larger raise in tariffs. The parameters we obtain are summarized in Table \ref{table:p19}. While we see small variations in the weight of the inputs from China and ROW, and the variance of demand. The key change is observed in the elasticity of inputs, $\sigma$, since it is calibrated to match the 3.23 percentage point decline in imports from China over sales observed from 2018 to 2024, given the rise in tariffs of $19$ percentage points. In this case, the elasticity of substitution equals 3.97, which is lower than the 4.5 used before. The larger change in the price of inputs from China allows for a smaller elasticity. 

\begin{center}
\resizebox{410pt}{!}{
\begin{threeparttable}
\caption{Parameters}
\begin{tabular}{ l c c | c c c  }
\multicolumn{6}{l}{Quarterly model, $T = 90$}	\\
\multicolumn{6}{l}{ } 		\\
\multicolumn{6}{l}{\textbf{Panel A. Calibrated parameters}	}	\\
Parameter					& 					& 	Value 	&  Moment			& Model	& Data 	 \\
\hline 
\hline														
Weight inputs China  		& $\theta^f$		& $0.084$		& inputs China / sales in 2018 		& $8.89\%$ 	& $8.89\%$  		\\
Weight inputs ROW 			& $\theta^c$		& $0.440$		& inputs ROW / sales in 2018 		& $27.75\%$ 	& $27.74\%$  		\\
Elasticity inputs  			& $\sigma$		& $3.97$		& $\Delta_{18-23}$ inputs China / sales	& $-3.24\%$ 	& $-3.23\%$  		\\
Variance of demand 			& $\sigma_{\nu}$	& $0.623$		&input inventory / sales in 2018	& $28.86\%$ 	& 	$28.86\%$  	\\
            			& \multicolumn{5}{l}{$\nu_j \sim \mathcal{LN}(0, \sigma_{\nu})$   } \\
                            \end{tabular}
\label{table:p19}
     \end{threeparttable}
     }
\end{center}

For this tariff change, we need a 30-day additional delay for the foreign inputs to match the observed input inventory over sales. Delays rise since inputs are less substitutable, thus we need a larger change in the standard deviation of delivery times to match the inventory data. Table \ref{table:t19} shows the results across the different stationary distributions. The rise in a 30-day delay and an increase in tariffs for inputs from China of $19$ percentage point leads to a drop in output of $8.5\%$ and an increase in prices of $2.1\%$. Inventories are influenced by two opposing factors. On one hand, the increase in tariffs prompts firms to shift away from Chinese inputs, which have the longest delivery times, reducing the need for inventories. On the other hand, as delays grow, firms are compelled to hold more inventories per unit used. The impact of delays outweighs the substitution effect, resulting in an $11.2\%$ increase in total inventories relative to sales. Isolating the effects of delays, we show that output drops by $2.4\%$ and the final good prices increase by $0.4\%$. In this case, inventories rise $17\%$. 
     
\resizebox{480pt}{!}{
\begin{threeparttable}
 \captionof{table}{Quantify costs of tariffs and delays with a $19$ percentage point increase}
\begin{tabular}{l | c c c c c c c }
    & (1) & (2) & (3) & (4) & (5) & (6) & (7) \\
			& \textbf{2018}			& \textbf{Tariffs}			& \textbf{Delays}		& \textbf{2024}		& \textbf{Tariffs vs}			& \textbf{Delays vs}		& \textbf{2024 vs}		 \\
                     & \textbf{Benchmark} &  \textbf{$\uparrow \tau^c$} & \textbf{$\uparrow$ delay} &  \textbf{$\uparrow \tau^c$ + delay} & \textbf{2018} & \textbf{2018} & \textbf{2018} \\
\hline
\hline
        Tariff             & $\tau^c p^c = 0.84$ 	& $\tau^c p^c = 1.0$ 	& $\tau^c p^c = 0.84$ & $\tau^c p^c = 1.0$ & & &  \\
         Change delays         &  $\Delta$ delay = 0	& $\Delta$ delay = 0	& $\Delta$ delay = 30 & $\Delta$ delay = 30  & & & \\
        Output & 0.919 & 0.860 & 0.897 & 0.841 & -6.5\% & -2.4\% & -8.5\%  \\ 
        Prices & 1.345 & 1.369 & 1.351 & 1.374 & 1.8\% & 0.4\% & 2.1\%  \\ 
        Inputs China/sales & 0.088 & 0.055 & 0.087 & 0.055 & -37.3\% & -1.7\% & -37.5\%  \\ 
        Inventories China/sales & 0.079 & 0.050 & 0.089 & 0.059 & -36.8\% & 13.6\% & -24.3\%  \\ 
        Inputs ROW/sales & 0.277 & 0.292 & 0.273 & 0.288 & 5.4\% & -1.6\% & 3.7\%  \\ 
        Inventories ROW/sales & 0.210 & 0.222 & 0.248 & 0.262 & 5.7\% & 18.3\% & 24.8\%  \\ 
        Inputs domestic/sales & 0.368 & 0.387 & 0.374 & 0.392 & 5.2\% & 1.6\% & 6.6\%  \\ 
        Inventories/sales & 0.288 & 0.271 & 0.337 & 0.321 & -5.9\% & 17.0\% & 11.5\% \\ 
    \end{tabular}
\label{table:t19}
\begin{tablenotes}
\item\footnotesize{Columns 1 to 4 in the table report the average of the stationary distribution for different variables. Column one represents the benchmark economy of 2018 with the low initial tariffs and no additional delays. Column two isolates the effects of tariffs, and computes the steady state average of an economy with the raise in tariffs and low delays. Column three shows an economy with high delays and the initially low tariffs. Then, column four shows an economy with both the rise in tariffs and delays. The last three columns compares the economies relatively to the benchmark 2018 economy to show the effects of tariffs, in column 5, the effect of delays, column 6, and both changes in column 7.}
\end{tablenotes}
     \end{threeparttable}
     }